\documentclass[twocolumn]{aastex631}
\usepackage{CJKutf8}
\usepackage{amsmath}

\shorttitle{The magnetic field in colliding filaments G202.3+2.5}
\shortauthors{Gu et al.}
\begin{document}
\begin{CJK}{UTF8}{gbsn}
\title{The Magnetic Field in Colliding Filaments G202.3+2.5}

\author[0000-0002-2826-1902]{Qi-Lao Gu (顾琦烙)}\thanks{E-mail:qlgu@shao.ac.cn}
\affiliation{Shanghai Astronomical Observatory, Chinese Academy of Sciences \\
No.80 Nandan Road, Xuhui, Shanghai 200030, People's Republic of China}

\author[0000-0002-5286-2564]{Tie Liu (刘铁)}\thanks{E-mail:liutie@shao.ac.cn}
\affiliation{Shanghai Astronomical Observatory, Chinese Academy of Sciences \\
No.80 Nandan Road, Xuhui, Shanghai 200030, People's Republic of China}

\author[0000-0001-8077-7095]{Pak Shing Li}
\affiliation{Shanghai Astronomical Observatory, Chinese Academy of Sciences \\
No.80 Nandan Road, Xuhui, Shanghai 200030, People's Republic of China}

\author[0000-0003-3540-8746]{Zhi-Qiang Shen (沈志强)}\thanks{zshen@shao.ac.cn}
\affiliation{Shanghai Astronomical Observatory, Chinese Academy of Sciences \\
No.80 Nandan Road, Xuhui, Shanghai 200030, People's Republic of China}

\author[0000-0001-8315-4248]{Xunchuan Liu (刘训川)}
\affiliation{Shanghai Astronomical Observatory, Chinese Academy of Sciences \\
No.80 Nandan Road, Xuhui, Shanghai 200030, People's Republic of China}

\author[0000-0002-4774-2998]{Junhao Liu(刘峻豪)}
\affiliation{National Astronomical Observatory of Japan \\
2-21-1 Osawa, Mitaka, Tokyo, 181-8588, Japan}

\author[0000-0003-2619-9305]{Xing Lu (吕行)}
\affiliation{Shanghai Astronomical Observatory, Chinese Academy of Sciences \\
No.80 Nandan Road, Xuhui, Shanghai 200030, People's Republic of China}

\author{Julien Montillaud}
\affiliation{Université de Franche-Comté, CNRS, Institut UTINAM, OSU THETA, F-25000 Besan\c{c}on, France}

\author{Sihan Jiao (焦斯汗)}
\affiliation{National Astronomical Observatories, Chinese Academy of Sciences\\
A20 Datun Road, Chaoyang, Beijing 100101, People's Republic of China}

\author[0000-0002-5809-4834]{Mika Juvela}
\affiliation{Department of Physics, PO box 64, FI-00014, University of Helsinki, Finland}

\author[0000-0002-6529-202X]{Mark G. Rawlings}
\affiliation{Gemini Observatory/NSF’s NOIRLab, 670 N. A’ohoku Place, Hilo, Hawai’i, 96720, USA}

\author[0000-0003-2384-6589]{Qizhou Zhang}
\affiliation{Center for Astrophysics | Harvard \& Smithsonian\\
60 Garden Street, Cambridge, MA 02138, USA}

\author[0000-0003-2777-5861]{Patrick Koch}
\affiliation{Academia Sinica Institute of Astronomy and Astrophysics\\
No. 1, Section 4, Roosevelt Road, Taipei 10617, Taiwan (R.O.C.)}

\author{Isabelle Ristorcelli}
\affiliation{IRAP, Universit\'{e} de Toulouse, CNRS \\
9 avenue du Colonel Roche, BP 44346, 31028 Toulouse Cedex 4, France}

\author{Jean-S\'{e}bastien Carriere}
\affiliation{IRAP, Universit\'{e} de Toulouse, CNRS \\
9 avenue du Colonel Roche, BP 44346, 31028 Toulouse Cedex 4, France}

\author{David Eden}
\affiliation{Armagh Observatory and Planetarium, College Hill, Armagh, BT61 9DB, UK}

\author[0000-0003-4659-1742]{Zhiyuan Ren (任致远)}
\affiliation{National Astronomical Observatories, Chinese Academy of Sciences\\
A20 Datun Road, Chaoyang, Beijing 100101, People's Republic of China}

\author[0000-0002-8149-8546]{Ken'ichi Tatematsu}
\affiliation{Nobeyama Radio Observatory, National Astronomical Observatory of Japan,
National Institutes of Natural Sciences\\
Nobeyama, Minamimaki, Minamisaku, Nagano 384-1305, Japan}
\affiliation{Astronomical Science Program,
Graduate Institute for Advanced Studies, SOKENDAI\\
2-21-1 Osawa, Mitaka, Tokyo 181-8588, Japan}

\author{Naomi Hirano}
\affiliation{Academia Sinica Institute of Astronomy and Astrophysics\\
No. 1, Section 4, Roosevelt Road, Taipei 10617, Taiwan (R.O.C.)}

\author[0000-0003-4506-3171]{Qiu-yi Luo (罗秋怡)}
\affiliation{Shanghai Astronomical Observatory, Chinese Academy of Sciences \\
No.80 Nandan Road, Xuhui, Shanghai 200030, People's Republic of China}
\affiliation{School of Astronomy and Space Sciences, University of Chinese Academy of Sciences \\
No.19A Yuquan Road, Beijing 100049, Peopleʼs Republic of China}

\author[0000-0001-7573-0145]{Xiaofeng Mai (麦晓枫)}
\affiliation{Shanghai Astronomical Observatory, Chinese Academy of Sciences \\
No.80 Nandan Road, Xuhui, Shanghai 200030, People's Republic of China}
\affiliation{School of Astronomy and Space Sciences, University of Chinese Academy of Sciences \\
No.19A Yuquan Road, Beijing 100049, Peopleʼs Republic of China}

\author{Namitha Issac}
\affiliation{Shanghai Astronomical Observatory, Chinese Academy of Sciences \\
No.80 Nandan Road, Xuhui, Shanghai 200030, People's Republic of China}

%% automatically takes care of all commas and "and"s between authors' names.


\begin{abstract}
We observe the magnetic field morphology towards a nearby star-forming filamentary cloud, G202.3+2.5, by the JCMT/POL-2 850 $\mu$m thermal dust polarization observation with an angular resolution of 14.4" ($\sim$0.053 pc). The average magnetic field orientation is found to be perpendicular to the filaments while showing different behaviors in the four subregions, suggesting various effects from filaments' collision in these subregions. With the kinematics obtained by ${\rm N_2H}^+$ observation by IRAM, we estimate the plane-of-sky (POS) magnetic field strength by two methods, the classical Davis-Chandrasekhar-Fermi (DCF) method and the angular dispersion function (ADF) method, ${\rm B}_{\rm pos,dcf}$ and ${\rm B}_{\rm pos,adf}$ are $\sim90 \mu \rm{G}$ and $\sim53 \mu \rm{G}$. We study the relative importance between the gravity (G), magnetic field (B) and turbulence (T) in the four subregions, find $G > T > B$, $G \geq T > B$, $G \sim T > B$ and $T > G > B$ in the north tail, west trunk, south root and east wing, respectively. In addition, we investigate the projection effect on the DCF and ADF methods based on a similar simulation case and find the 3D magnetic field strength may be underestimated by a factor of $\sim3$ if applying the widely-used statistical ${\rm B}_{\rm pos}$-to-${\rm B}_{\rm 3D}$ factor when using DCF or ADF method, which may further underestimate/overestimate related parameters.

\end{abstract}

%% You will be asked to select these concepts during the submission process

%% and \citet command to identify citations.  The citations are

\section{Introduction} \label{sec:intro}
The Magnetic field (B-field) is one of the major players in the star formation process along with gravity and turbulence \citep{Crutcher2012, Brandenburg2013, Seifried&Walch2015, Offner2018, Pattle2023}. With the increasing polarization observations toward molecular clouds in the last decades, the B-field is found to be dynamically important in different stages of the star formation process over a large range of spatial scales: (a) on the scale of a few to ten pc, the molecular clouds tend to be either parallel with or perpendicular to the global average B-field orientation \citep[e.g.][]{Soler2013, Li2013, Gu&Li2019}, suggesting that the B-field is strong enough to guide gas flow; (b) on the scale of a few tenths to a pc, the B-field is found to interact with the gas motion from filaments to cores and provide support against gravitational collapse \citep[e.g.][]{Seifried&Walch2015, Tang2019}. 

Filaments are ubiquitous in molecular clouds and are found to be a significant stage of the star formation process \citep{Myers2009, Arzoumanian2011, LiuBb2012a, Andre2014, Lin2017, Lu2018}, among which velocity coherent fiber-like structures are identified \citep[e.g.][]{Li2012, LiuBb2012b, Hacar2013, Henshaw2016} with chains of dense cores embedded \citep[e.g.][]{Zhang2009, Tafalla&Hacar2015, Zhang2015}, suggesting the fragmentation from fibers to chains of cores. Recent large-scale MHD simulations of the formation of filamentary clouds\citep[e.g.][]{Klassen2017, LiPS2018, Li&Klein2019} found that B-field is generally piercing through the elongated direction of filaments or fibers. \cite{LiuT2018} reported a pinched B-field in IRDC G035.39-0.33 at a distance of 2.9 kpc, hinting at an accretion flow along the filaments. \cite{Busquet2013} reported a network of filaments in the infrared dark cloud G14.225-0.506 at a distance of 2.3 kpc. B-fields probed by polarized dust emissions in the near-infrared \citep{Santos2016} and sub-millimeter wavelengths \citep{AnezLopez2020} are nearly perpendicular to the main axis of the filaments. Velocity gradients are identified along the filaments, suggestive of mass accretion \citep{Chen2019}. Here we report on a much more nearby source, G202.3+2.5, that provides more evidence that accretion flows can be traced via pinched B-field surrounding chains of cores.

\begin{figure*}[ht!]
\centering
\includegraphics[scale=1]{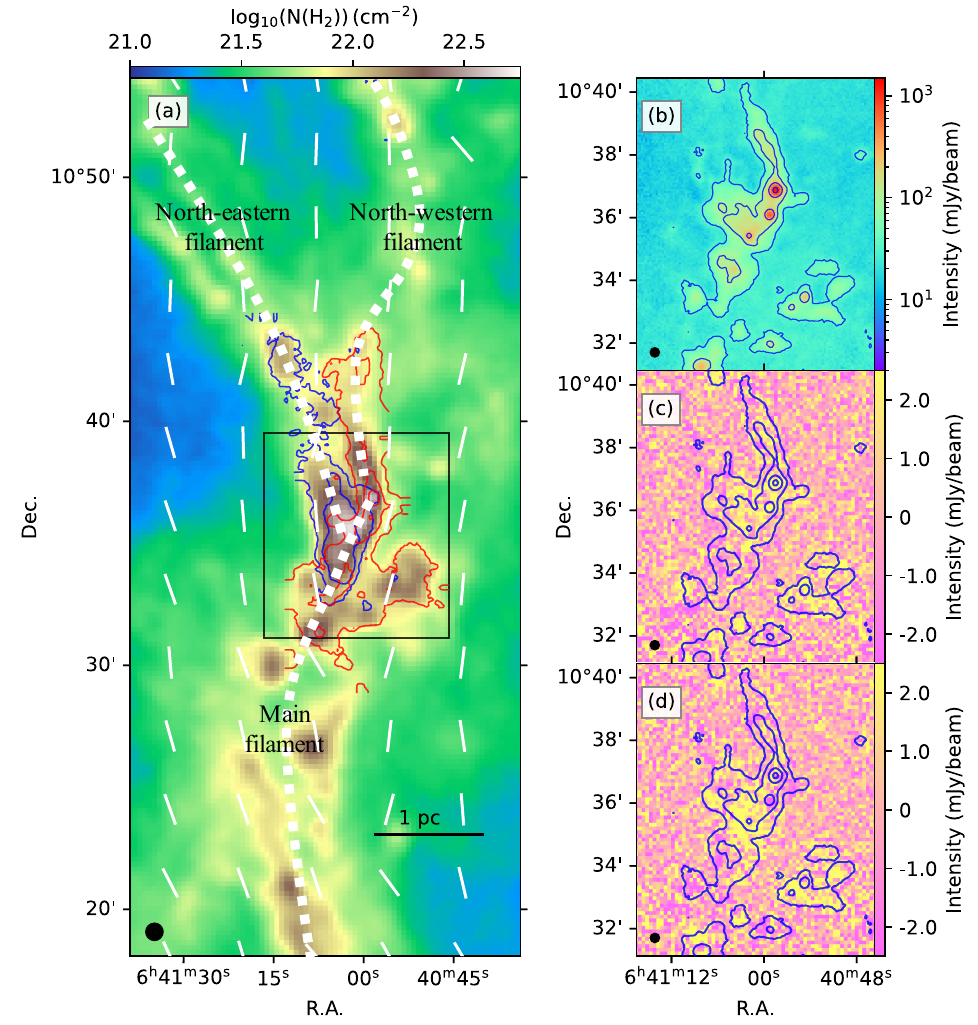}
\caption{\em{\textbf{(a)} Column density of molecular hydrogen in G202.3+2.5 with a resolution of 38.5$\arcsec$ (adopted form \cite{Montillaud2015}), blue and red contours show the CS (J=2-1) moment 0 data from the IRAM data (adapted form \cite{Montillaud2019b}), which are integrated from velocity channel 4.2 km/s to 6.6 km/s and 7.8 km/s to 9.6 km/s, respectively. The curved dashed white lines sketch the structures of the main filament, north-eastern filament, and north-western filament. White segments represent the B-field orientations inferred from Planck 353 GHz polarization data and the black rectangle denotes the region shown in (b-d), a 1 pc scale bar is shown in the lower right corner; \textbf{(b-d)} JCMT/POL-2 850 $\mu$m Stokes I, Q and U maps of G202.3+2.5, respectively, and contours show the intensity of Stokes I at the levels of [35, 100, 300 and 1000] mJy/beam with an average measured rms level of 10 mJy/beam. The beams are shown in the left corners.}}
\label{fig:iqu}
\end{figure*}

G202.3+2.5 is a nearby star-forming filamentary structure that is located at the edge of the massive ($3.7\times10^4 M_{\odot}$) molecular complex Monoceros OB 1 at a distance of 723 pc \citep{cantat-gaudin_gaia_2018}. With observations of the TRAO (Taeduk Radio Astronomy Observatory) 14-m and IRAM (Institut de Radio Astronomie Millim\'{e}trique) 30-m telescopes, \cite{Montillaud2019b} found that G202.3+2.5 is a complex and ramified structure (Figure~\ref{fig:iqu} (a)), among which two filaments (hereafter the north-western and north-eastern filaments) join into another larger and denser one (hereafter the main filament). Red and blue CS (J=2-1) contours show two well-separated velocity components, standing for the northernmost part of the main filament and the southernmost part of the north-eastern filament, respectively. Based on the typical chemical age of  ${\rm N_2H}^+$, \cite{Montillaud2019b} proposed that the junction region (marked as a black rectangle) is the result of a collision that started $\sim 10^5$\, yr ago between the two components and which might trigger the formation of the cores in the main filament (listed in Table~\ref{tb1}).

However, details of the present state and the history of G202.3+2.5, especially the role the B-field plays in the process, are poorly known, \cite{Carriere2022} and \cite{Alina2022} found that  G202.3+2.5 and its host complex Monoceros OB 1 both tend to be parallel with the B-field by $4.8\arcmin$-resolution ($\sim$ 1 pc at a distance of 723 pc) Planck 353 GHz polarization data, suggesting B-field remains dragged and is providing support against fragmentation at the scale larger than 1 pc, but how the B-field behaves in the denser region is still unknown. Here we present JCMT/POL-2 850 $\mu$m dust polarization observations to investigate the interaction between the B-field and star formation process in G202.3+2.5 from a deeper view at a scale of 0.05 pc.

This paper is organized as follows: in Section~\ref{sec2}, we describe the observations and data reduction; we show the results in Section~\ref{sec3}; in Section~\ref{sec4}, we discuss our result and further analyses with magnetohydrodynamics (MHD) simulation; finally, we summarize our findings in Section~\ref{sec5}.

\section{Observations}\label{sec2}
The POS B-field orientation is derived from polarized 850 $\mu$m continuum observations through SCUBA-2/POL-2 on JCMT. G202.3+2.5 was observed 37 times from 2019 February to 2019 March and 2019 December to 2020 January (project code: M19AP032 and M19BP019; PI: Tie Liu) using SCUBA-2/POL-2 DAISY mapping mode \citep{Holland2013, Friberg2016, Friberg2018} under Band 1 weather conditions ($\tau_{225GHz}$ < 0.05, where $\tau_{225GHz}$ is the atmospheric opacity at 225 GHz), providing a total integration time of $\sim$ 21.2 hr. The effective beam size is $14.4\arcsec$ at 850 $\mu$m \citep{Mairs2021}, equivalent to $\sim$ 0.050 pc at a distance of 723 pc. A single POL-2 observation consists of around 40 minutes of observing time and produces a $12\arcmin$ diameter output map, of which the central $3\arcmin$ has approximately uniform noise, and the central $6\arcmin$ has a useful level of coverage \citep{Friberg2016}.

The data were reduced using a STARLINK package named SMURF \citep{Chapin2013, Currie2014}, which is specifically developed for submillimeter data reduction from JCMT. The main processes are as follows\footnote{The details can be found in \url{http://starlink.eao.hawaii.edu/docs/sc22.htx/sc22.html}}: 

(1) The \textit{calcqu} command converted raw bolometer timestreams into separate Stokes \textit{I}, \textit{Q} and \textit{U} timestreams, and the \textit{makemap} routine in script \textit{pol2map} generated individual \textit{I} maps from \textit{I} timestreams and co-added them to produce an initial reference \textit{I} map; 

(2) The \textit{pol2map} command was rerun with the initial \textit{I} map to create two masks, ASTMASK (which defines background regions in the reference \textit{I} map) and PCAMASK (which defines source regions to exclude when creating the background model within the \textit{makemap} routine). 

(3) The \textit{pol2map} command was rerun again with ASTMASK and PCAMASK to create improved final \textit{I} maps, which were co-added to the final \textit{I} map. With the same parameters and masks, \textit{Q} and \textit{U} maps, along with the corresponding variance maps and polarization catalog were created as well. The final \textit{I}, \textit{Q}, \textit{U} maps and polarization catalog were gridded to 7$\arcsec$/pixel for a Nyquist sampling\footnote{Typically, these maps are gridded to 12$\arcsec$/pixel for optimized noise and S/N \citep[e.g.][]{Kwon2018, Eswaraiah2021}, we choose 7$\arcsec$/pixel for a balance between relative good S/N and enough B-field segments for applying DCF method.}.

The final \textit{I}, \textit{Q} and \textit{U} maps are in units of pW. We converted them to the unit of Jy beam$^{-1}$ by applying the flux conversion factor (FCF) of 668 Jy beam$^{-1}$ pW$^{-1}$. This value is 1.35 times larger than the standard SCUBA-2 850 $\mu$m FCF 495 Jy beam$^{-1}$ pW$^{-1}$  because of the flux loss from POL-2 \citep{Mairs2021}. Figure \ref{fig:iqu} (b-d) shows the Stokes \textit{I}\footnote{The Stokes \textit{I} map shown in Figure~\ref{fig:iqu} (b) is a combination of SCUBA-2 850 $\mu$m and deconvolved Planck 353 GHz using J-comb algorithm from \cite{Jiao2022}, the details is described in Section~\ref{sec:3.2.1}}, \textit{Q}, and \textit{U} maps of these JCMT/POL-2 observations. The rms levels of the background regions are $\sim$10 mJy/beam in \textit{I} and $\sim$1.5 mJy/beam in \textit{Q} and \textit{U}.  

Due to the uncertainties of \textit{Q} (\textit{$\delta$Q}) and \textit{U} (\textit{$\delta$U}), and the polarized intensity (\textit{P}) and polarized percentage (\textit{p}) are defined as positive values, the measured polarized intensities are biased toward larger values \citep{Vaillancourt2006}. The debiased \textit{P} and its uncertainty (\textit{$\delta$P}) are calculated from, 
\begin{equation}
P = \sqrt{Q^2+U^2-0.5(\delta Q^2+\delta U^2)},
\label{eq:pi}
\end{equation}
and
\begin{equation}
\delta P = \sqrt{\frac{Q^2 \delta Q^2+U^2 \delta U^2}{Q^2+U^2}}.
\label{eq:dpi}
\end{equation}
Therefore the debiased \textit{p} and its uncertainty (\textit{$\delta$p}) are derived by,
\begin{equation}
p = \frac{P}{I},
\label{eq:p}
\end{equation}
and
\begin{equation}
\delta p = \sqrt{\frac{\delta P^2}{I^2}+\frac{P^2\delta I^2}{I^4}},
\label{eq:dp}
\end{equation}
where \textit{$\delta$I} is the uncertainty of \textit{I}. Finally, polarization angle (\textit{$\theta$}) and its uncertainty (\textit{$\delta \theta$}) \citep{Naghizaden-khouei1993} are estimated by 
\begin{equation}
\theta = \frac{1}{2}tan^{-1}(\frac{U}{Q}),
\label{eq:theta}
\end{equation}
and
\begin{equation}
\delta \theta = \frac{1}{2P}\sqrt{Q^2\delta U^2+U^2\delta Q^2}.
\label{eq:dtheta}
\end{equation}

We note that there are many debiasing methods for polarization data \citep[e.g.][]{Montier2015a, Montier2015b}, but in this work, we use \textit{P} for pseudo-vectors selection only, thus the effect of our debiasing method is minimal on our results. The derived B-field angles are not determined from the absolute calibration, but the relative values of \textit{Q} and \textit{U}\citep{Pattle2017}.

%\footnote{\url{https://www.eaobservatory.org/jcmt/2019/08/new-ip-models-for-pol2-data/}}
\begin{figure*}[ht!]
\centering\includegraphics[scale=0.85]{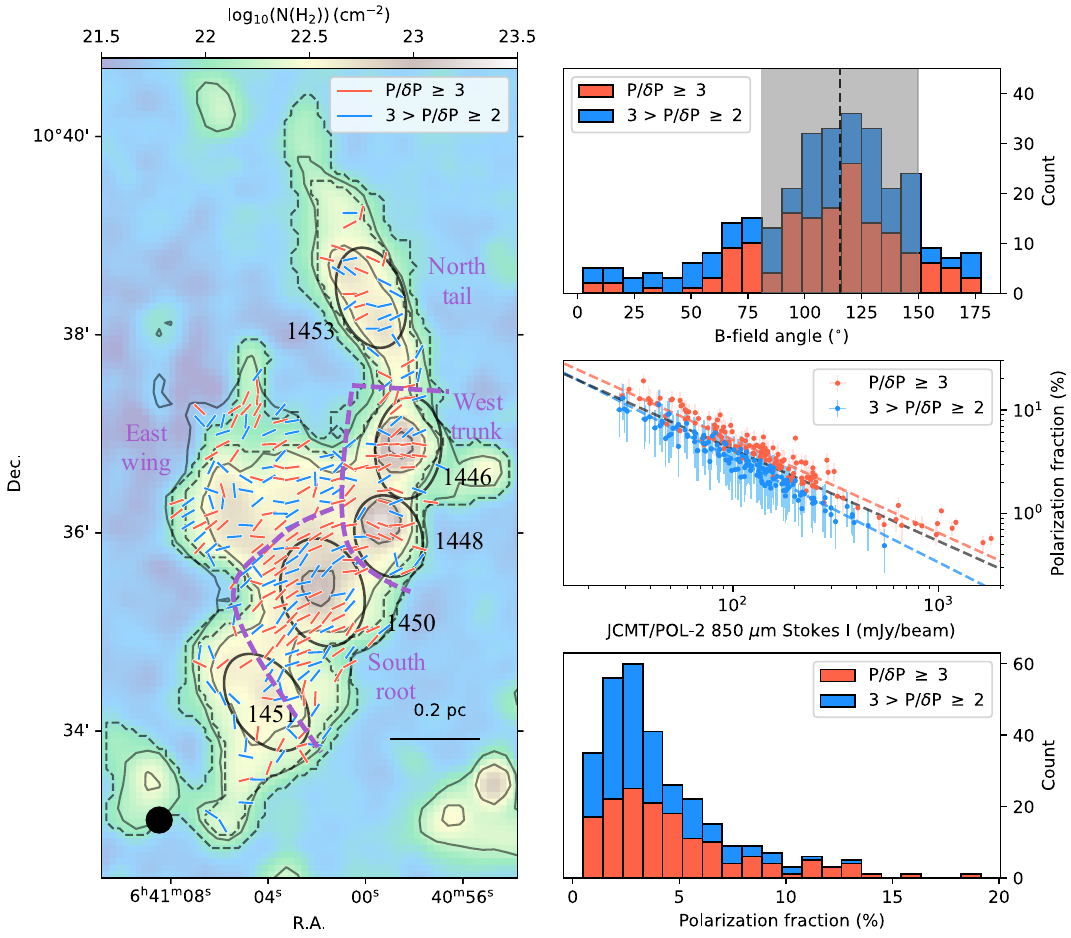}
\caption{\em{\textbf{Left panel}: Observed B-field of G202.3+2.5 region. The color map is the $N(H_2)$ of molecular hydrogen with a resolution of 18$\arcsec$, the beam is shown in the lower left corner. Dashed purple lines divided G202.3+2.5 into four subregions, the east wing, north tail, west trunk, and south root, respectively. Black dashed contour marks the $N(H_2)$ level of $1.5\times10^{22}\,{\rm cm}^{-2}$, solid contours show the intensity of Stokes I at the levels of [35, 100, 300 and 1000] mJy/beam. Red and blue segments represent JCMT/POL-2 B-field orientations with S/N levels of $P/\delta P\geq 3$ and $3\geq P/\delta P\geq2$, respectively. Black ellipses mark the location of cores listed in Table~\ref{tb1}. A 0.2 pc scale bar is shown in the lower right corner and the 18$\arcsec$ beam is shown in the lower left corner. \textbf{Upper right panel}: Histogram of B-field orientations inferred from JCMT/POL-2 with different $P/\delta P$ levels, the dashed line represents the average B-field orientation of all the segments and the grey region marks the standard deviation range. \textbf{Middle right panel}: Polarization fraction vs. Stokes I. Blue and red colors represent data with different $P/\delta P$ levels, dashed red and blue lines are the power-law fits, respectively. The dashed black line is the power-law fit for all data. \textbf{Lower right panel}: Histogram of polarization fraction with different $P/\delta P$ levels.}}
\label{fig:b_cd}
\end{figure*}

\begin{deluxetable*}{cccrrcccccc}
\tablenum{1}
\tablecaption{Parameters of cores inside G202.3+2.5}
\tablewidth{0pt}
\tablehead{
\colhead{Core ID} & \colhead{R.A.} & \colhead{Dec.} & \colhead{P.A.} & \colhead{$B_{POL2}$} & \colhead{Mass} & \colhead{$N_{\rm H_2}$} & \colhead{R} & \colhead{$n({\rm H}_2)$} & \colhead{Class} & \colhead{State}\\
\colhead{} & \colhead{h m s} & \colhead{d m s} & \colhead{($^{\circ}$)} & \colhead{($^{\circ}$)} & \colhead{($M_{\odot}$)} & \colhead{($10^{22} {\rm cm}^{-2}$)} & \colhead{pc} & \colhead{($10^{5} {\rm cm}^{-3}$)} & \colhead{} & \colhead{}
}
\decimalcolnumbers
\startdata
1446 & 6 40 58.2432 & 10 36 51.48 & -5 & $98.2^{+16.6}_{-14.7}$ & 21.6 & 4.40 & 0.092 & 1.21 & I & 4\\
1448 & 6 40 59.0568 & 10 35 58.092 & 20.9 & $88.2^{+21.5}_{-20.4}$ & 15.3 & 3.79 & 0.083 & 1.05 & I & 4\\
1450 & 6 41 1.7664 & 10 35 24.036 & 16.5 & $134.2^{+13.9}_{-11.1}$ & 51.7 & 7.89 & 0.106 & 1.76 & U/I & 0\\
1451 & 6 41 4.0608 & 10 34 17.436 & 36 & $142.6^{+30.6}_{-23.7}$ & 16.3 & 2.94 & 0.097 & 0.77 & U/I & 3\\
1453 & 6 40 59.7744 & 10 38 22.38 & 17.9 & $91.1^{+19.1}_{-25.0}$ & 22.3 & 4.49 & 0.092 & 1.24 & U/I & 2\\
\enddata
\caption{The columns of the upper table give parameters of cores: (1) ID number; (2) the right ascension; (3) the declination; (4) position angle; (5) average B-field orientation angle inferred from JCMT/POL-2 850$\mu$m data, superscript and subscript show the interquartile range (IQR) with setting average as $Q_2$ (the second quartile); (6) mass; (7) column density; (8) radius that calculated by $R=\sqrt{a\times b}$, where a and b are the FWHM major and minor axes at 350 $\mu$m that were adopted from \cite{Montillaud2015}; (9) volume density, was calculated from (7) and (8); (10) class of embedded YSO, I for Class I and U/I for undetermined or irrelevant; (11) final state of evolution, 0 for undetermined or irrelevant, 1 for candidate starless source, 2 for reliable starless source, 3 for candidate protostellar source, 4 for confirmed protostellar source. (1), (2), (3), (4), (6), (7), (10) are adopted from \cite{Montillaud2015}, (11) is adopted from \cite{Montillaud2019a}.}
\label{tb1}
\end{deluxetable*}

\section{Results}\label{sec3}
\subsection{Dust polarization properties}\label{sec3.1}
Based on the grain alignment theory that the shortest axis of dust grains tends to align with the B-field \citep{Lazarian2003}, we can derive the projected POS B-field orientation by rotating the observed polarization pseudo-vectors by $90^{\circ}$. The polarization pseudo-vectors used here are selected with a criterion of $I/\delta I\geq10$, $P/\delta P\geq2$, and $\delta p \leq 5\%$. The inferred B-field segments are shown in the left panel of Figure~\ref{fig:b_cd}, overlaid on the $N({\rm H}_2)$ column density map with an average orientation of $113\pm33^{\circ}$, roughly showing a perpendicular alignment with the main structure of G202.3+2.5. 

As shown in the middle right panel of Figure~\ref{fig:b_cd}, there is a trend of decreasing polarization fraction with increasing dust emission intensity (i.e. the depolarization effect), a power law with index -0.89 is fitted to all data (dashed black line, the power law index is -0.90 and -1.00 for data with $P/\delta P\geq3$ and $3\geq P/\delta P\geq2$, respectively). The upper right panel is the histogram of inferred B-field angles with different $P/\delta P$ levels, with an average value of $113\pm33^{\circ}$\footnote{All the position angles shown in this paper follow the IAU-recommended convention of measuring angles from the north towards the east.} marked as a dashed grey line and the average values of $P/\delta P\geq3$ and $3\geq P/\delta P\geq2$ are $112\pm31^{\circ}$ and $117\pm37^{\circ}$. The lower right panel shows the distribution of the polarization fraction. The distribution peaks at $\sim$3\% with a tail extending to $\sim$10\% - 20\%. The average, standard deviation and median of polarization fraction are 4.2\%, 3.0\%, and 3.3\% for the whole data. For the data with $P/\delta P\geq3$, the corresponding values are 4.8\%, 3.5\%, and 3.8\%, and for data with $3\geq P/\delta P\geq2$, the values are 3.6\%, 2.4\%, 2.9\%. As shown in the right panels, segments with $P/\delta P\geq3$ and $3\geq P/\delta P\geq2$
have similar B-field angle and polarization fraction distributions, thus we include them together for the following analysis and discussion to increase the statistics. 

\subsection{B-field strength of G202.3+2.5}\label{sec:3.2}

Because of the lack of direct ways to measure the B-field strength in molecular clouds, the DCF method \citep{Davis1951, CF1953} is widely used to give a rough estimate. The DCF method relies on an assumption that there is a prominent mean (uniform/ordered) B-field component ($B_0$) with a perturbation ($B_t$, turbulent B-field) caused by turbulence, which assumes there is equipartition between the turbulent magnetic energy ($E_{B_t}=B_t^{\phantom{t}2}V/(8\pi)$ \footnote{All of the magnetic-related equations and terms in this work are in cgs units unless otherwise noted.}) and the turbulent kinetic energy ($E_K=\rho V\sigma_{v,\perp}^{\phantom{v,\perp}2}/2$, where V is the volume, $\sigma_{v,\perp}$ is the transverse turbulent velocity dispersion, and $\rho$ is the gas density). In addition, assuming the turbulence is isotropic, $\sigma_{v,\perp}$ equals the LOS velocity dispersion $\sigma_v$. Thus the strength of the POS turbulent B-field, $B_{\rm pos,t}$, could be derived from $B_{\rm pos,t}=\sqrt{4\pi\rho}\sigma_v$, and the strength of the POS uniform B-field could be estimated from 
\begin{equation}
B_{\rm pos} \simeq f_{\rm DCF}\sqrt{4\pi\rho}\frac{\sigma_v}{B_{\rm pos,t}/B_{\rm pos}},
\label{eq:dcf}
\end{equation}
where $f_{\rm DCF}$ is a correction factor, and the $B_{\rm pos,t}/B_{\rm pos}$ could be roughly estimated from the angular dispersion between the local B-field orientation and the mean one ($\sigma_{\theta}$). 

Recently, studies of DCF methods show different ways to quantify $B_{pos,t}/B_{pos}$ more accurately \citep[e.g.][]{Heitsch2001, Falceta2008, CY2016, LiuJH2021, Chen2022}. Here we propose two of them to estimate the B-field strength: the modified classical DCF method and the calibrated angular dispersion function method \citep[hereafter the ADF method, a modified DCF method raised by][]{Hildebrand2009, Houde2009, Houde2016}. 

\subsubsection{Estimation of gas density and LOS velocity dispersion}\label{sec:3.2.1}

\begin{figure*}[ht!]
\centering\includegraphics[scale=1]{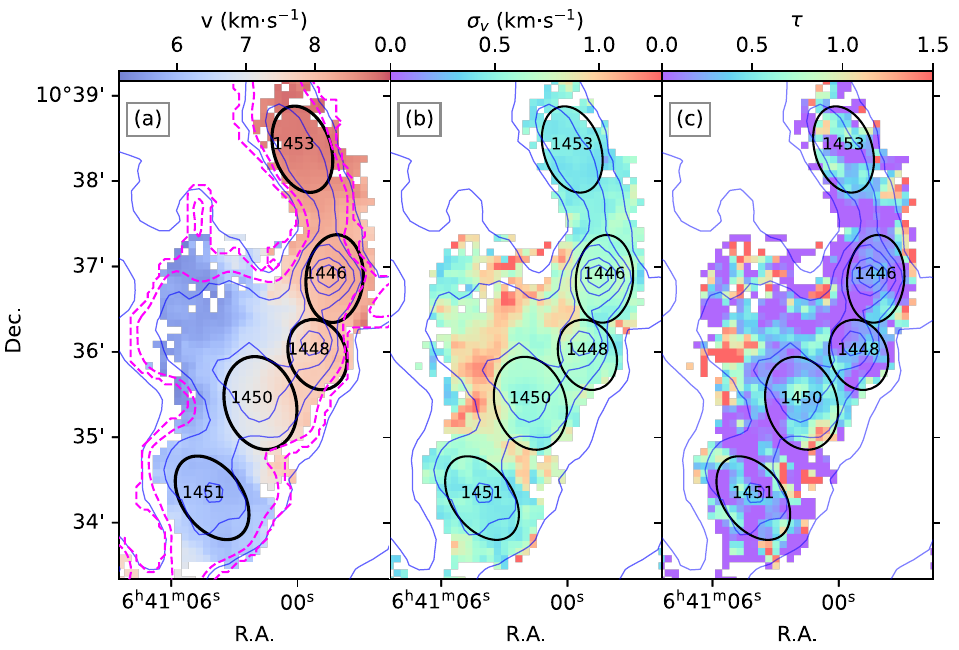}
\caption{\em{\textbf{Panel (a)}: Centroid map of the isolated hyperfine component of $N_2H^+$ in G202.3+2.5, blue contours show the JCMT/POL-2 850 $\mu$m Stokes I with levels of [20, 100, 200, 500, 1000] mJy/beam, dashed magenta contours show the $N(H_2)$ with levels of [8,15]$\times10^{21}\,cm^{-2}$; \textbf{Panel (b)}: $N_2H^+$ Velocity dispersion; \textbf{Panel (c)}: Optical depth from hyperfine structure fitting. Black ellipses in each panel mark the location of cores listed in Table~\ref{tb1}.}}
\label{fig:n2hp}
\end{figure*}

Before estimating the gas density ($\rho$), we apply the J-comb algorithm \citep{Jiao2022} to get a column density map with higher resolution (left panel of Figure~\ref{fig:b_cd}) than the one adopted from \cite{Montillaud2015} (Figure~\ref{fig:iqu} (a)) based on our JCMT 850 $\mu$m data as follows:

(1) Data selection. We retrieved the level 2.5 processed, archival \textit{Herschel} images that were taken at 70/160 $\mu$m using the Photodetector Array Camera and Spectrometer \citep[PACS, obsID: 1342228372,][]{Poglitsch2010} and at 250/350/500 $\mu$m using the Spectral and Photometric Imaging REceiver \citep[SPIRE, obsID: 1342228342,][]{Griffin2010}. We used the \textit{Herschel} SPIRE extended emission products from the database.

(2) Derive combined Stokes I map. We extrapolated an 850 $\mu$m flux map from SED of \textit{Herschel} SPIRE data and used this map as a model image to deconvolve the 353 GHz Planck map \citep{Planck2011} with Lucy-Richardson algorithm \citep{Lucy1974}. The obtained deconvolved Planck map has an angular resolution close to the SPIRE 500 $\mu$m map and preserves the flux level of the initial Planck map. The deconvolved map was then combined with the JCMT/SCUBA-2 850 $\mu$m Stokes I map in the Fourier domain using J-comb algorithm \citep{Jiao2022} to derive the combined Stokes I map (Figure~\ref{fig:iqu} (b)).

(3) SED fitting. We smoothed \textit{Hershel} 70/160/250 $\mu$m and the combined JCMT/SCUBA-2 450/850 $\mu$m images to a common angular resolution of the largest telescope beam, and all maps were re-gridded to have the same pixel size. We weighted the data points by the measured noise level in the least-squares fits. As a modified black-body assumption, the flux density ${\rm S}_{\nu}$ at frequency $\nu$ is given by 
\begin{equation}
{\rm S}_{\nu} =\Omega_m {\rm B}_{\nu}(T_{\rm dust})(1−e^{−\tau_{\nu}}),   
\end{equation}
where $\Omega_m$ is the solid angle, $B_{\nu}(T_{\rm dust})$ is the Planck function at temperature ${\rm T}_{\rm dust}$, and the column density N$({\rm H}_2)$ then can be approximated by 
\begin{equation}
{\rm N}({\rm H}_2) = \tau_{\nu} /\kappa_{\nu} \mu m_H 
\end{equation}
where $\kappa_{\nu} = 0.1\,{\rm cm}^2\,{\rm g}^{-1} (\nu/1000\,\rm{GHz})^{\beta}$ is the adopted dust opacity assuming a gas-to-dust ratio of 100 and an opacity index $\beta$ of 2 \citep{Hildebrand1983, Beckwith1990}\footnote{We use the opacity index of 2 to match the column density map adopted from \cite{Montillaud2015} (shown in Figure~\ref{fig:iqu} using the same value) and larger scale result in \cite{Alina2022}.}, $\mu = 2.8$ is the mean molecular weight and $m_{\rm H}$ is the atomic mass of hydrogen.

% Dust temperature $T_{dust}$ and $N({\rm H}_2)$ were computed by combining three \textit{Hershel} SPIRE \citep[The Spectral and Photometric Imaging REceiver,][]{Griffin2010} maps at 250, 350, and 500 $\mu$m, which were convolved to a resolution of 38.5". $N({\rm H}_2)$ was fitted by
% \begin{equation}
% N({\rm H}_2) = \frac{I_{\nu}}{B_{\nu}(T_{\rm dust})\kappa_{\nu}\mu m_{\rm H}},
% \label{eq:n_h2}
% \end{equation}
% % spectral energy distribution (SED) was fitted by 
% % \begin{equation}
% % I_{\nu} = B_{\nu}(T_{dust})\nu^{\beta}
% % \label{eq:i_nu}
% % \end{equation} 
% in each pixel, where $I_{\nu}$ is the dust emission intensity at frequency $\nu$ and $B_{\nu}(T_{\rm dust})$ is the Planck function at $T_{\rm dust}$, $\kappa_{\nu} = 0.1\,{\rm cm}^2\,{\rm g}^{-1} (\nu/1000\,\rm{GHz})^{\beta}$ is the adopted dust opacity assuming a gas-to-dust ratio of 100 and an opacity index $\beta$ of 2 \citep{Hildebrand1983, Beckwith1990}, $\mu = 2.8$ is the mean molecular weight and $m_{\rm H}$ is the atomic mass of hydrogen.
\begin{figure*}[ht!]
\centering\includegraphics[scale=0.5]{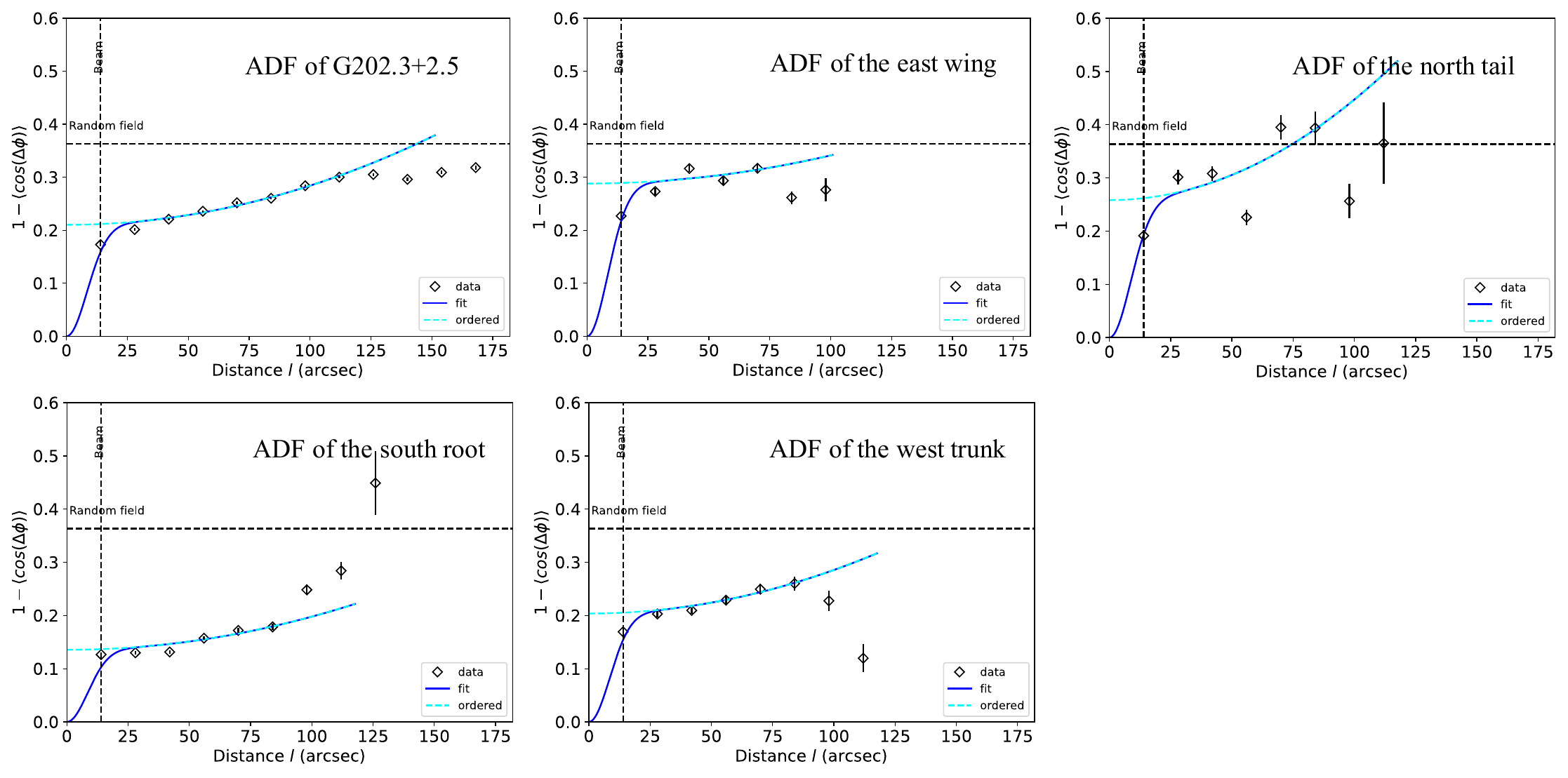}
\caption{\em{ADFs of G202.3+2.5 and its subregions. An effective depth is calculated based on the dataset before ADF fitting, and only dots that are at a distance smaller than the effective depth are fitted.}}
\label{fig:adf}
\end{figure*}
The new column density map (left panel of Figure~\ref{fig:b_cd}) has a resolution of 18$\arcsec$, and the mass above ${\rm N}({\rm H}_2)$ a contour of $1.5\times10^{22}\,{\rm cm}^{-2}$ (inner dashed contour in left panel of Figure~\ref{fig:b_cd}) is $274\,M_{\odot}$ and above ${\rm N}({\rm H}_2)$ a contour of $8.0\times10^{21}\,{\rm cm}^{-2}$ (outer dashed contour in left panel of Figure~\ref{fig:b_cd}) is $330\,M_{\odot}$\footnote{The unadjacent structures shown in the left panel of Figure~\ref{fig:b_cd} (e.g. the one in the lower right corner) are not counted in.}. Considering G202.3+2.5 is found to be the junction region of two colliding filaments with lengths of $\sim1.3$ pc and radius of $\sim0.1$\,pc \citep{Montillaud2019b}, we assume G202.3+2.5 as two cylinders with a length of $\sim1.3$ pc and a radius of $\sim0.1$ pc to estimate the volume. According to the two values of mass are estimated above, $\rho$ is then derived as $2.27\times10^{-19}\,{\rm g \, cm}^{-3}$ and $2.73\times10^{-19}\,{\rm g \, cm}^{-3}$, respectively, and we take the average value $(2.50\pm0.23)\times10^{-19}\,{\rm g \, cm}^{-3}$ (the corresponding volume density is $(5.34\pm0.49)\times10^4\,{\rm cm}^{-3}$) for the following analysis.

As for LOS velocity dispersion, $\sigma_v$, we extracted kinematic information from IRAM 30-m ${\rm N_2H}^+ (J=1-0)$ observations with a resolution of 27.8$\arcsec$, which is adopted from \cite{Montillaud2019b}. As shown in Figure~\ref{fig:n2hp} (a), detected ${\rm N_2H}^+ (J=1-0)$ region matches well with $[0.8,1.5]\times10^{22}{\rm cm}^{-2} {\rm N}({\rm H}_2)$ contours and Stokes I contours, suggesting ${\rm N_2H}^+ (J=2-1)$ traces the densest region of G202.3+2.5. The FWHM line width, $\Delta v$, is derived from hyperfine structure line fitting, and the non-thermal component $\sigma_v$ (Figure~\ref{fig:n2hp} b) is then derived from 
\begin{equation}
\frac{\Delta v}{\sqrt{8ln(2)}} = \sqrt{c_s^{\phantom{s}2}+\sigma_v^{\phantom{v}2}},
\label{eq:dv_nt}
\end{equation}
where $c_s=\sqrt{\frac{k_BT}{m_{\rm N_2H^+}}}$ is the isothermal sound speed and $T$ is the dust 
temperature (the average is 13.0 K with a standard deviation of 1.0 K) adopted from \cite{Montillaud2019b}, and the average $\sigma_v$ of G202.3+2.5 is $0.66\pm0.21$ km/s.

\subsubsection{The classical DCF method}\label{sec:3.2.2}
A $f_{\rm DCF}$ of 0.5 is widely used based on the simulation result from \cite{Ostriker2001}, and Equation~\ref{eq:dcf} could be modified as
\begin{equation}
B_{\rm pos} \simeq 0.5\sqrt{4\pi\rho}\frac{\sigma_v}{\sigma_{\theta}}.
\end{equation}
However, this formula is only applicable on a scale larger than 1 pc with $\sigma_{\theta} < 25^{\circ}$ \citep{Heitsch2001, LiuJH2022, Kate2022}. In order to remove the small-$\sigma_{\theta}$ restriction, \cite{Li2022} replaced $\sigma_{\theta}$ by $\tan(\sigma_{\theta})$ \citep[cf.][]{Heitsch2001, Falceta2008}. Thus, the modified classical DCF method could be written as 
\begin{equation}
B_{\rm pos,dcf} \simeq 0.5\sqrt{4\pi\rho}\frac{\sigma_v}{\tan(\sigma_{\theta})},
\label{eq:dcf_classical}
\end{equation}
and the $B_{\rm pos,dcf}$ of G202.3+2.5 is derived as $90\pm37 \,\mu{\rm G}$. Considering $B_{\rm pos} = B \sin(\phi)$ for an angle of $\phi$ between LOS and $\textbf{B}$, the strength of total 3-D B-field, $B_{\rm 3D,dcf}$, can be estimated statistically as $114\pm 47\, \mu{\rm G}$ by $B_{\rm 3D}=\frac{4}{\pi}B_{\rm pos}$ statistically \citep{Crutcher2004}.

\subsubsection{The calibrated ADF method}
Alternatively, $B_{\rm pos,t}/B_{\rm pos}$ can be represented as $\left(\frac{\langle B_t^{\phantom{t}2}\rangle}{\langle B^2\rangle}\right)^{1/2}$, the ratio of turbulent-to-total B-field strength on POS, which is determined by the structure function of polarization angles \citep{Falceta2008}. \cite{Hildebrand2009, Houde2009} further expanded the method to account for larger-scale field structures and LOS effects, respectively. Further, with numerical simulation, \cite{LiuJH2021} calibrated the ADF method and found it correctly accounts for the ordered B-field structure and beam smoothing. Here, we use the calibrated ADF method \citep{LiuJH2021, LiuJH2022}, and the routine is outlined as follows:

(1) Derive the ADF that accounts for the ordered field and the POS turbulent correlation effect from
\begin{equation}\label{eq:adf}
% \begin{split}
1-\langle\cos[\Delta \Phi(l)]\rangle \simeq\frac{\langle B_t^{\phantom{t}2}\rangle}{\langle B^2\rangle}\times(1-e^{-l^2/2(l_{\delta}^2+2W^2)})+a'_2l^2,
% \end{split}
\end{equation}
where $\Delta \Phi(l)$ is the angular difference of two B-field line segments separated by a distance of $l$, $l_{\delta}$ is the turbulent correlation length for local turbulent B-field, $W$ is the standard deviation of the Gaussian beam and $a'_2l^2$ is the first term of Taylor expansion of the ordered component of ADF. For the case that $l\gg l_{\delta}$ and $l\gg W$, Equation~\ref{eq:adf} can be simplified to  
\begin{equation}\label{eq:adf_simplified}
% \begin{split}
1-\langle\cos[\Delta \Phi(l)]\rangle \simeq\frac{\langle B_t^{\phantom{t}2}\rangle}{\langle B^2\rangle}+a'_2l^2.
% \end{split}
\end{equation}

(2) Fit the large-scale portion of the ADF (Equation~\ref{eq:adf_simplified}) to estimate $\frac{\langle B_t^{\phantom{t}2}\rangle}{\langle B^2\rangle}$, which is the intercept of the fitted ordered ADF as shown in Figure~\ref{fig:adf}, the reduced function in Equation~\ref{eq:adf_simplified} constructed with the best-fit results are shown as dashed cyan curves. The best-fit $(\frac{\langle B_t^{\phantom{t}2}\rangle}{\langle B^2\rangle})^{1/2}$ of G202.3+2.5 and its subregions are listed in Table~\ref{tb_subregions}.
 
(3) Estimate the total POS B-field strength, $B_{\rm pos,adf}$, by 
\begin{equation}
B_{\rm pos,adf} \simeq 0.21\sqrt{4\pi\rho}\sigma_v(\frac{\langle B_t^{\phantom{t}2}\rangle}{\langle B^2\rangle})^{-1/2},
\label{eq:adf_tot}
\end{equation} 
where $f_{\rm DCF}$ is taken as 0.21\citep[with 45\% uncertainty at scales between 0.1 and 1 pc,][]{LiuJH2021}, which gives $B_{\rm pos,adf}\simeq52.7\pm 24.5 \, \mu{\rm G}$. Also, the strength of total 3-D B-field, $B_{\rm 3D,adf}$, could be estimated as $67.1\pm 31.2 \,\mu{\rm G}$ by the same process described in Section~\ref{sec:3.2.2}.

\begin{figure}[ht!]
\centering\includegraphics[scale=0.5]{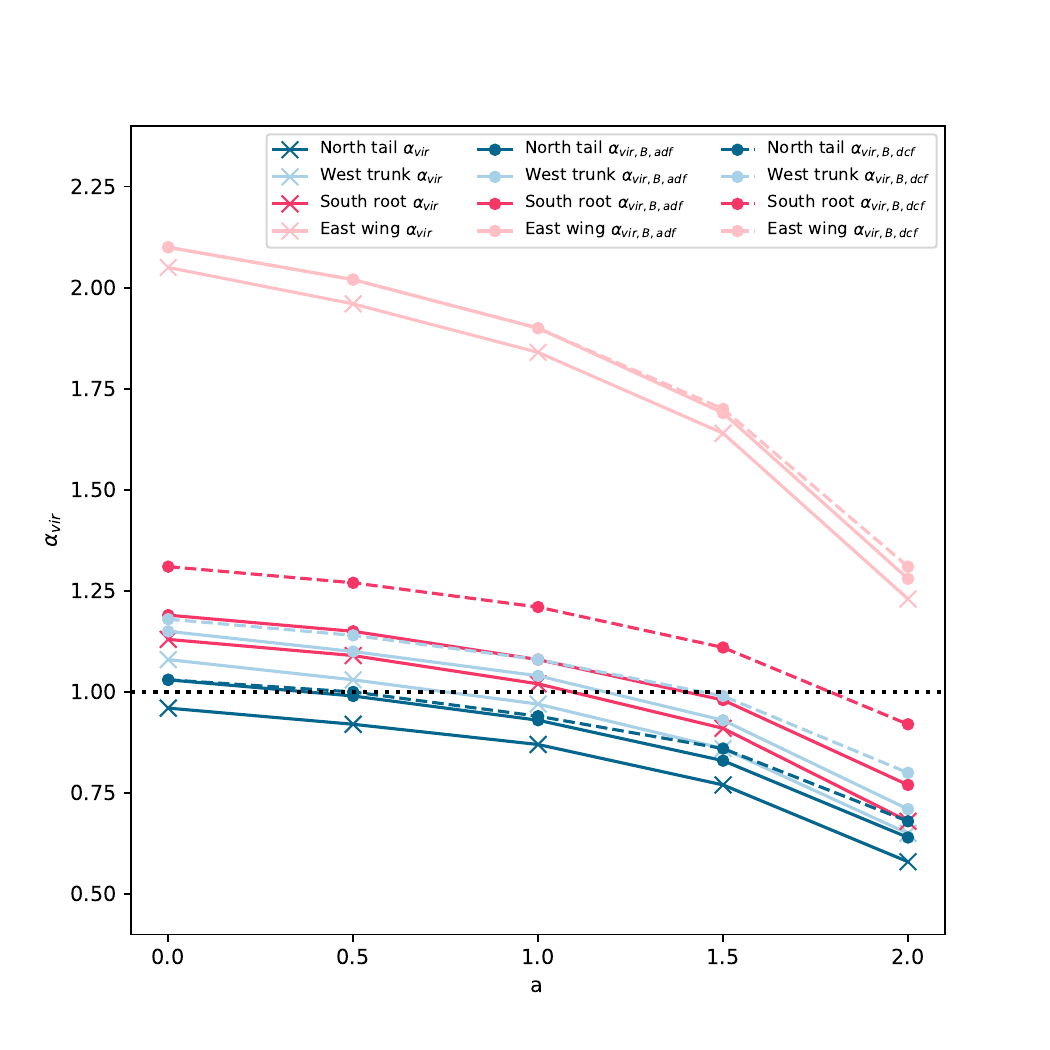}
\caption{\em{Virial parameter $\alpha_{\rm vir, B}$ as a function of index $a$ of the radial density profile.}} Solid and dashed lines with dot markers show $\alpha_{\rm vir,B,adf}$ and $\alpha_{\rm vir,B,dcf}$, respectively. Solid lines with cross markers show $\alpha_{\rm vir}$. An $\alpha_{\rm vir}=1$ line is shown in a dotted black line.
\label{fig:alpha}
\end{figure}

\begin{figure*}[ht!]
\centering\includegraphics[scale=0.6]{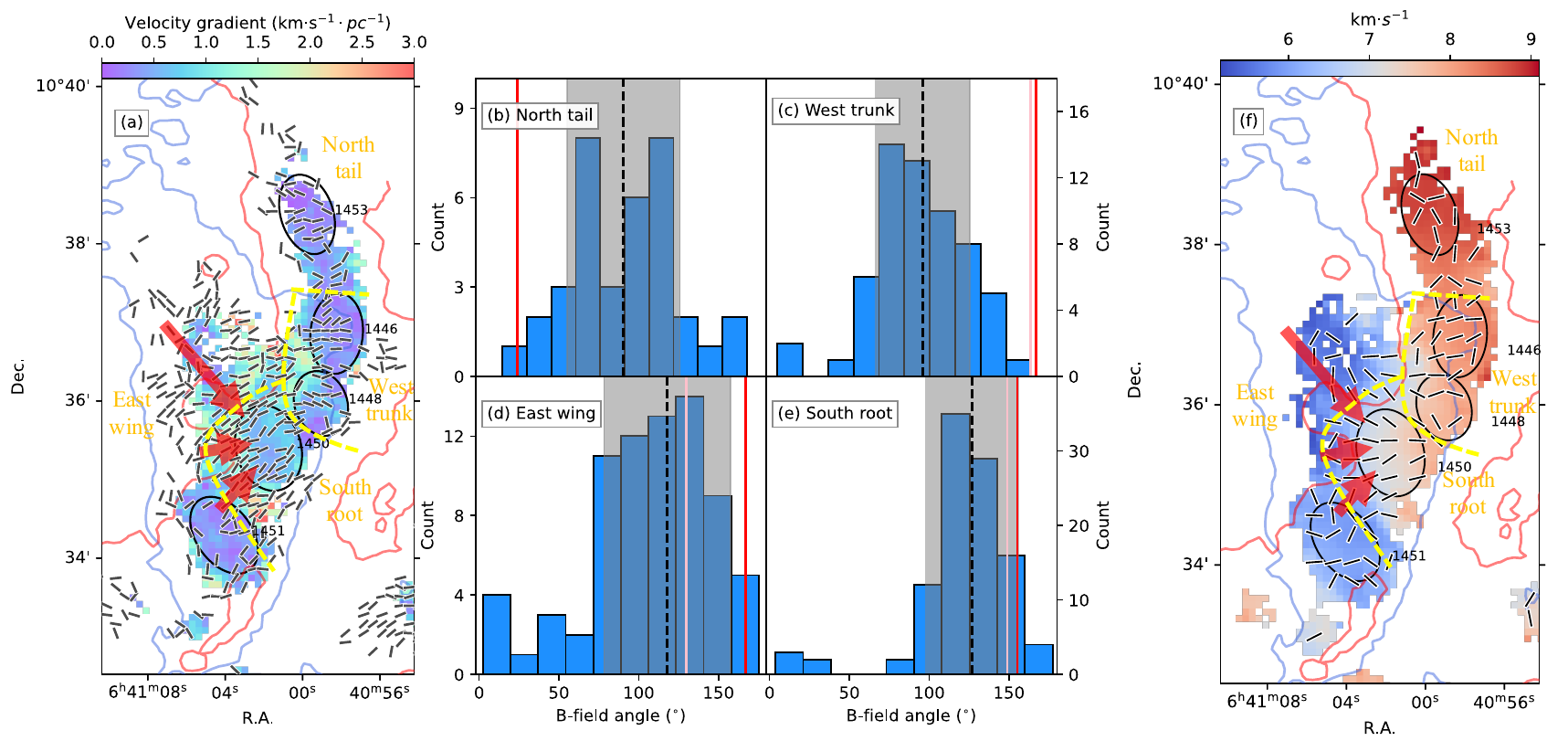}
\caption{\em{\textbf{Panel (a)}: Relative $N_2H^+$ velocity gradient overlaid by the B-field. Red and blue are the outermost contours of CS (J=2-1) shown in panel (a) of Figure~\ref{fig:iqu}, black ellipses mark the locations of cores. \textbf{Panel (b) - (e)}: Distributions of B-field orientations in the four subregions. Solid red and pink lines mark the density elongation direction derived by auto-correlation function based on ${\rm N}{(\rm H}_2$ contours of $1.5\times10^{22} {\rm cm}^{-2}$ and $8.0\times10^{21} {\rm cm}^{-2}$, respectively. Dashed black lines show the average orientation of B-field, and grey regions mark the standard deviation range. \textbf{Panel (f)}: Centroid map of the isolated hyperfine component of N$_2$H$^{+}$ in G202.3+2.5 overlaid by the velocity gradient direction segments. In panel (a) and (f), dashed yellow lines divided G202.3+2.5 into north tail, west trunk, east wing, and south root, respectively. Red arrows indicate the possible gas flow directions.}}
\label{fig:b_vg_components}
\end{figure*}

\section{Discussion}\label{sec4}
As shown in the left panel of Figure~\ref{fig:b_cd}, the B-field morphology is different in different regions of G202.3+2.5, thus we divide G202.3+2.5 into four regions to do further analysis: north tail, east wing, west trunk, and south root\footnote{The B-field segments in the southmost part of G202.3+2.5 are discrete with those in the south root, thus we haven't counted them into the south root in for a more precise calculation of B-field strength.}. In the north tail, the B-field is roughly perpendicular to the dense core 1453 whose long axis is parallel with the filament, but with larger dispersion. In the west trunk, the B-field is perpendicular to the filament and shows a typical 'hourglass' shape around the dense core 1446. The situation is different in the other two regions. In the south root, where the southernmost tip of the north-eastern filament has already merged into the main filament, the B-field is parallel with the filament, but perpendicular to the core 1450. As for the east wing, the frontline of collision, the B-field is significantly disturbed, which is reflected in the orientation of the B-field changing from parallel with the north-eastern filament to perpendicular to it.

\subsection{Gravity vs. B-field vs. turbulence}\label{sec:GBT}
To quantitatively compare the relative importance of B-field ($B$), turbulence ($T$), and gravity ($G$) towards subregions, we have calculated some key parameters as follows based on the B-field strength calculated above, the results are listed in Table~\ref{tb_subregions}. As the B-field strength has been estimated by two methods, these parameters also have two versions.

\subsubsection{Alfv\'{e}n Mach numbers}
The relative importance between the turbulence and B-field can be parameterized by the Alfv\'{e}n Mach number
\begin{equation}
\mathcal{M}_A = \sigma_{v,3D}/v_A,
\label{eq:ma}
\end{equation}
where $\sigma_{v,3D}$ is derived from $\sigma_{v,3D}=\sqrt{3}\sigma_v$ based on the isotropic turbulence assumption, $v_A = B_{\rm 3D}/\sqrt{4\pi\rho}$ is the velocity of the Alfv\'{e}n wave. It is worth noting that considering Equation~\ref{eq:dcf_classical} and Equation~\ref{eq:adf_tot}, and assuming $B_{\rm 3D} = \frac{4}{\pi}B_{\rm pos}$ \citep{Crutcher2004}, $\mathcal{M}_A$ could be written as a function of $\left(\frac{\langle B_t^{\phantom{t}2}\rangle}{\langle B^2\rangle}\right)^{1/2}$ (Equation~\ref{eq:ma_adf}) or $\sigma_{\theta}$ (Equation~\ref{eq:ma_dcf}).

\begin{equation}
\mathcal{M}_{A,adf} = \sigma_{v,3D}/v_{A,adf} \approx 6.48\left(\frac{\langle B_t^{\phantom{t}2}\rangle}{\langle B^2\rangle}\right)^{1/2},
\label{eq:ma_adf}
\end{equation}

\begin{equation}
\mathcal{M}_{A,dcf} = \sigma_{v,3D}/v_{A,dcf} \approx 2.72\tan(\sigma_{\theta}),
\label{eq:ma_dcf}
\end{equation}

In all subregions, $\mathcal{M}_{A}$ is larger than 1, indicating a $T > B$ status. However, it is worth noting that as $v_A$ depends on $B_{\rm 3D}$, which relates to the the ${\rm B}_{\rm pos}$-to-${\rm B}_{\rm 3D}$ factor a lot. The ${\rm B}_{\rm pos}$-to-${\rm B}_{\rm 3D}$ factor applied here is derived statistically as $4/\pi$, may cause a large bias for a specific case, further discussion is shown in Section~\ref{sec:sim}.

\subsubsection{Virial parameters}
The virial parameter, $\alpha_{\rm vir}$, is used to estimate the relative importance of the kinetic support against gravity and is defined via

\begin{equation}
\alpha_{\rm vir}=\frac{2E_{\rm K}}{|E_{\rm G}|}=\frac{3(5-2a)\sigma^2{R}}{(3-a){GM}}=\frac{M_{\rm K}}{M},
\label{eq:vir}
\end{equation}
where $a$ is the index of radial density profile ($\rho\propto r^{-a}$), $\sigma_v$, $R$, G, $M$, and $M_K$ are velocity dispersion, radius, the gravitational constant, mass, and kinetic virial mass, respectively \citep{Bertoldi1992}. If taking the effects of the B-field into account, the magnetic virial parameter, $\alpha_{{\rm vir},B}$, can be calculated following \cite{Liu2020} via

\begin{equation}
\alpha_{{\rm vir},B}=\frac{M_{\rm K+B}}{M},
\label{eq:vir_b}
\end{equation}
where

\begin{equation}
M_{\rm K+B} = \sqrt{M_{\rm B}^{\phantom{B}2}+(M_{\rm K}/2)^2}+M_{\rm K}/2,
\label{eq:m_kb}
\end{equation}
and

\begin{equation}
M_{\rm B} = \frac{\pi R^2B}{\sqrt{\frac{3(3-a)}{2(5-2a)}\mu_0\pi G}}.
\label{eq:m_b}
\end{equation}
The effective radius could be estimated from $\rm R = \sqrt{A/\pi}$, where $A$ is the effective area above the column density of $1.5\times10^{21}\,{\rm cm}^{-2}$, then $\alpha_{{\rm vir},B, {\rm adf}}$ and $\alpha_{{\rm vir},B,{\rm dcf}}$ could be derived, respectively. 

Due to the limited resolution, the value of $a$ is hard to estimate, as $a=0$ and $a=2$ indicate a uniform density profile and a gravitationally bounded one, respectively, the real $a$ towards subregions should be in between, thus we calculate $\alpha_{vir}$ with $a\in[0,2]$ as shown in Figure~\ref{fig:alpha} to give an estimated region of $\alpha_{vir}$. Considering the west trunk and south root have higher ${\rm N}({\rm H}_2)$ and embed dense cores already, we speculate an $a$ between 1 and 2 in these two regions,  which gives $\alpha_{\rm vir,B}\sim1$ and $\alpha_{\rm vir}<1$ in the west trunk and $\alpha_{\rm vir,B}\geq1$ and $\alpha_{\rm vir}<1$ in the south root, indicating the statuses of $G\geq T$ and $G\sim T$, respectively. For the other two subregions, $\alpha_{\rm vir}$ and $\alpha_{\rm vir,B}$ are smaller than 1 in the north tail and larger than 1 in the east wing, suggesting status of $G > T$ and $T > G$, respectively.

\subsubsection{Mass-to-flux ratios}
We determine the ratio of mass to magnetic flux, $M/\Phi$, in units of the local magnetic stability critical parameter $\lambda$ \citep{Crutcher2004} to estimate the relative importance of the B-field and gravity:

\begin{equation}
\lambda = \frac{(M/\Phi)_{observed}}{(M/\Phi)_{critical}},
\label{eq:lambda1}
\end{equation}
where $(M/\Phi)_{observed}$ is the observed ratio given by

\begin{equation}
(\frac{M}{\Phi})_{observed} = \frac{\mu m_{\rm H} N({\rm H}_2)}{B_{\rm pos}},
\label{eq:mphi_obs}
\end{equation}
and $(M/\Phi)_{critical}$ is the critical ratio given by 

\begin{equation}
(\frac{M}{\Phi})_{critical} = \frac{1}{2\pi\sqrt{G}}.
\label{eq:mphi_cri}
\end{equation}
Thus, we estimated $\lambda$ by the relation in \cite{Crutcher2004}:

\begin{equation}
\lambda = 7.6\times10^{-21}\frac{N({\rm H}_2)}{B_{\rm pos}},
\label{eq:lambda2}
\end{equation}
where $N({\rm H}_2)$ is the column density in the unit of ${\rm cm}^{-2}$ and $B_{\rm pos}$ is POS total B-field strength in the unit of $\mu{\rm G}$. Value $\lambda<1$ indicates that the B-field is strong enough to support the gravitational collapse (magnetically subcritical) while $\lambda>1$ means that the B-field cannot prevent the gravitational collapse (magnetically supercritical). $\lambda$ in the host region of G202.3+2.5 (North-Main) is $0.95\pm0.41$ \citep{Alina2022}, indicating a trans-critical status at the scale larger than 1 pc, or (e.g.) the large-scale B-field is still relatively comparable to the gravity. As for G202.3+2.5, $B_{\rm pos,adf}$ and $B_{\rm pos,dcf}$ are $52.7\pm24.5\mu{\rm G}$ and $89.6\pm36.7\mu {\rm G}$, respectively, the average column density is $(3.18\pm0.47)\times10^{22}\,{\rm cm}^{-2}$, thus we find $\lambda_{\rm adf}\sim3.5\pm1.8$ and $\lambda_{\rm dcf}\sim2.1\pm 1.1$. \cite{Crutcher2004} proposed that the $(M/\Phi)_{observed}$ will be overestimated by up to a factor of 3 due to geometrical effects, if so it implies that $\lambda_{\rm adf}\sim1.2\pm 0.6$ and $\lambda_{\rm dcf}\sim0.7\pm 0.4$, respectively. However, since G202.3+2.5 appears to lie in or near the POS \citep{Montillaud2019b}, the factor of 3 is too large, thus, a $\lambda$ that is slightly larger than 1 is more convincing, suggesting that on the scale of the filament, G202.3+2.5 is slightly magnetically supercritical or trans-critical. Towards the four subregions, $\lambda$ is also larger than 1 even if excessively accounting for the geometrical effect, suggesting a magnetically supercritical status ($G > B$) on the scale of 0.05\,pc.

Therefore, we are able to conclude the relative importance of $G$, $B$ and $T$: in the north tail, $G > T > B$, where has a starless core 1453 with a mass of $22.3\,{\rm M}_{\odot}$; in the west trunk, $G \geq T > B$, where dense cores 1446 and 1448 with mass of $15-22\,{\rm M}_{\odot}$ have already formed and embedded Class I YSO; in the south root, $G\sim T > B$, where a massive core 1450 with a mass of $51.7\,{\rm M}_{\odot}$ forms and still accumulating mass from gas flows from the filaments' collision, which makes $T$ as large as $G$; in the east wing, $T > G > B$, where the turbulence dominates, and no dense cores have formed. 

\subsection{Compressed B-field in the collision filaments}\label{sec:CompressedB}

\begin{figure*}[ht!]
\centering\includegraphics[scale=0.8]{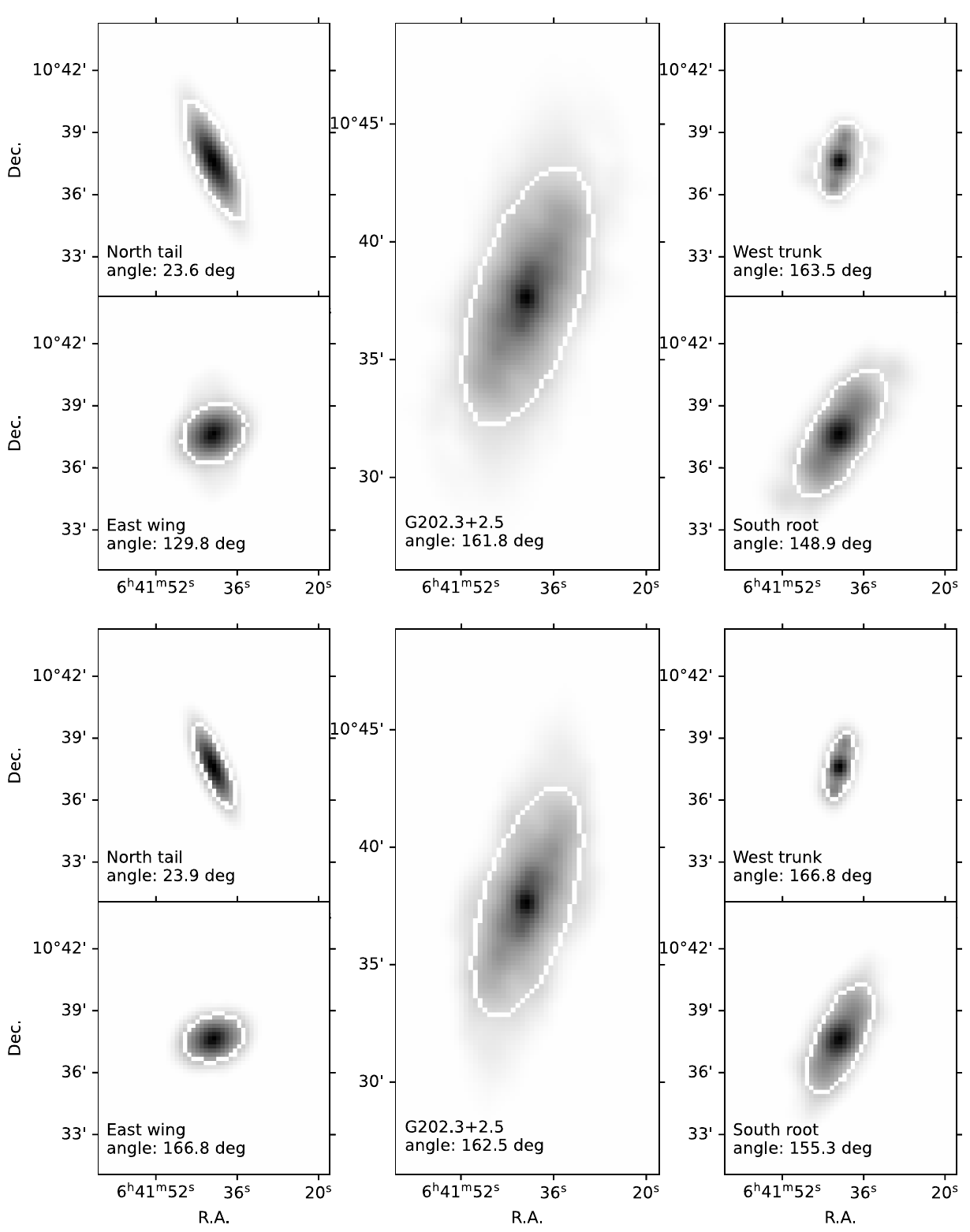}
\caption{\em{The autocorrelation maps of G202.3+2.5 and four subregions. Upper and lower panels show the result based on column density above contours level of $8.0\times10^{21}\,{\rm cm}^{-2}$ and $1.5\times10^{22}\,{\rm cm}^{-2}$, respectively. Background maps show the column density maps after applying the auto-correlation function, and white ellipses are fitted to estimate the elongation directions.}}
\label{fig:ac}
\end{figure*}

We apply the auto-correlation function to estimate the elongation direction of the column density contour (Figure~\ref{fig:ac}) to compare the relative orientation between the B-field and filament structure to study the interaction between the B-field and collision.

The north tail and west trunk have no overlap with the north-eastern filament, the velocity dispersion (Figure~\ref{fig:n2hp} (b)) and gradient (Figure~\ref{fig:b_vg_components} (a))\footnote{The gradient of LOS velocity (Figure~\ref{fig:b_vg_components} (a), $\partial v_z/\partial x, \partial v_z/\partial y$) at each pixel is calculated by taking the difference of the velocities 3 pixels away in both R.A. and Dec. directions, where the pixel size is 5$\arcsec$. It is worth noting that the calculated velocity gradient actually is the change of LOS velocity on the POS, as we cannot observe the POS component of turbulence and we have assumed isotropic turbulent motions when applying the DCF method, we treat this calculated gradient as the real gradient (the change of POS velocity on the POS).} are relatively small, and the B-field morphology is simple, suggesting that the B-field is not affected by the collision badly in these two regions. To be specific, in the north tail, which is the most diffuse subregion with an average ${\rm N}({\rm _H}_2)$ of $2.49\times10^{22}\,{\rm cm}^{-2}$, a bimodal distribution of the B-field angle is found (Figure~\ref{fig:b_vg_components} (b)), which results in a large $\sigma_{\theta}$ of $42\pm14^{\circ}$. The peak at $60^{\circ}$ comes from the outskirt B-field while the other at $110^{\circ}$ is contributed to the perpendicular B-field inside the protostellar core 1453. In the west trunk, the densest part of the main filament where two Class I sources (core 1446 and core 1448) are embedded, the B-field shows a perpendicular alignment with the structure, and a $\lambda_{\rm adf} = 4.8$ $(\lambda_{\rm dcf} = 3.0)$ indicates the magnetically supercritical status of the west trunk, thus, we suggest the gravity is dominating the B-field and drags the field lines. 

In Figure~\ref{fig:b_vg_components}, red arrows mark the possible direction of the gas flow traced by the velocity gradient, showing possible gas flow accumulation towards core 1450. Along the longest arrow, the B-field orientation changes rapidly, with a large velocity dispersion (Figure~\ref{fig:n2hp} (b)), indicating that the B-field has been disturbed significantly in the east wing. In contrast, in the south root, where the two filaments are overlapped in the LOS, the B-field strength is the strongest of the four subregions, and its average orientation is close to the one in the east wing, suggesting the B-field is compressed from the east wing to the south root by the collision. The orientation of the B-field is turned around to be parallel with the filament, and the transition is clear in the interface of the west trunk and south root, where the B-field orientation turns nearly $90^{\circ}$. The compression signature shown in the south root may hint that the gas flows toward the dense cores in return; similar behavior is also seen in distant IRDC \citep{LiuT2018} and simulation results \citep{LiPS2018}. 

On the scale larger than 1\,pc, \cite{Carriere2022} and \cite{Alina2022} found a roughly parallel cloud-field alignment towards the host region of G202.3+2.5 (North-Main) based on Planck data. In this work, we find a perpendicular cloud-field alignment towards G202.3+2.5 on the scale of 0.05\,pc, but in the south root, the denser region of G202.3+2.5, the alignment turns back to parallelism due to the compression from the gas flow caused by filaments' collision. \cite{Pillai2020} found a similar parallel-to-perpendicular-to-parallel transition with increasing density in a hub-filament system, Serpens South Cloud, where the perpendicular-to-parallel transition happens at $A_v\sim$20 mag ($N_{\rm H_2} \sim 4\times10^{22} {\rm cm}^{-2}$) is thought to be caused by merging gas flow reorienting the B-field. Though the south root has a simpler structure than Serpens South Cloud, the similar perpendicular-to-parallel transition at a similar density indicates the B-field can be reoriented by the gas flow to be parallel with the filament at a high-density region.

Based on the discussion above, we suggest the following scenario: in the east wing, the north-eastern filament dashes towards the main filament. The collision causes a large velocity gradient and dispersion and disturbs the B-field orientation, which results in a large B-field angle dispersion, but the B-field is still resisting the alignment with gas flow and is compressed in the frontline of the collision. While in the south root, gravity effects dominate, gas flow compresses the B-field into the direction perpendicular to the north-eastern filament which is also the elongation direction of the main filament, and the B-field strength increases due to the compression. The gas flow accumulates more mass onto the massive core 1450, which already has a mass of $51.7\,M_{\odot}$.

\begin{figure*}[ht!]
\centering\includegraphics[scale=0.65]{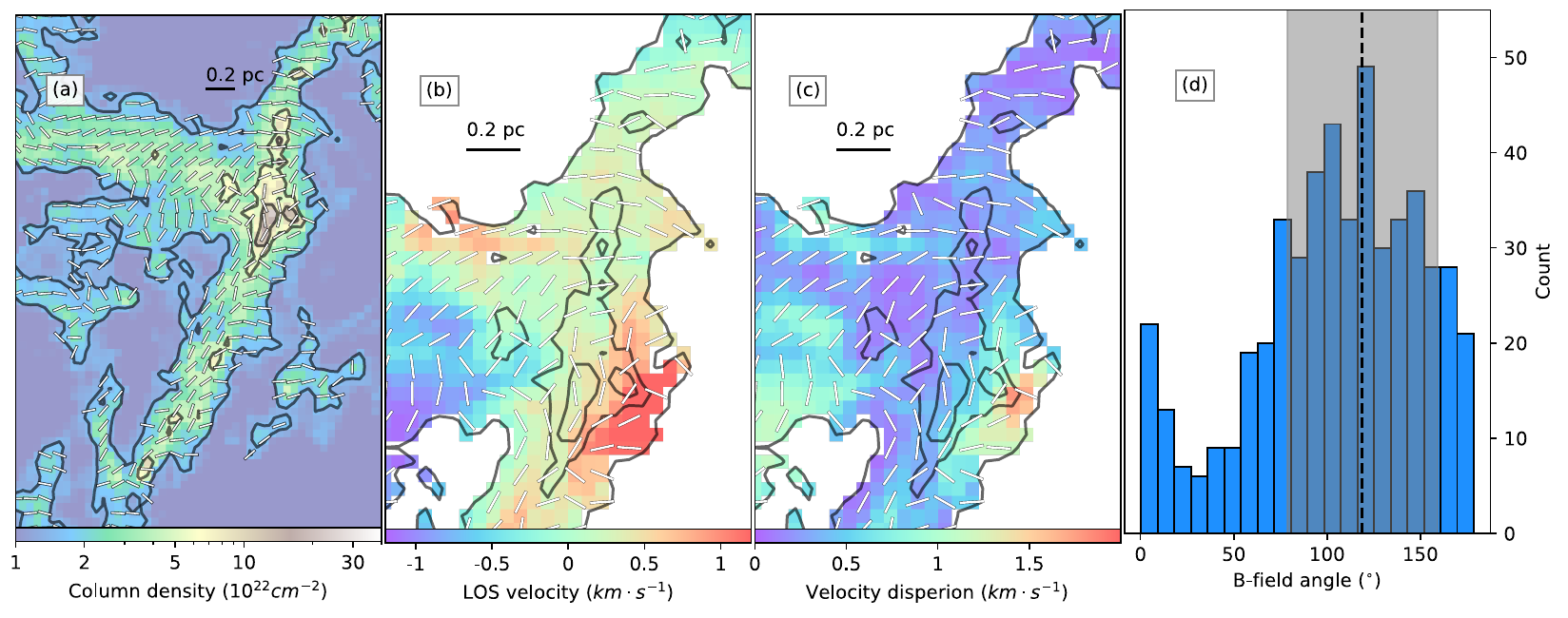}
\caption{\em{Analysis of an MHD simulation of colliding filaments. Contours in all panels show the column density with levels of [1.5,5,10]$\times10^{22}\,{\rm cm}^{-2}$, overlaid segments show the B-field orientations (plotted every $2\times2$ pixels for clarity). \textbf{Panel (a)}: Column density of simulated filamentary structures with a pixel size of 10000 au; \textbf{Panel (b)}: LOS velocity map of the same regions as panel (c); \textbf{Panel (c)}: Velocity dispersion of the same regions as panel (b); \textbf{Panel (d)}: Histogram of B-field orientations for pixels with a ${\rm N}({\rm H}_2)$ larger than $1.5\times10^{22}\,{\rm cm}^{-2}$. The dashed black line shows the average orientation, 118.8$^{\circ}$, and the gray region the standard deviation range $\pm40.3^{\circ}$.}}
\label{fig:sim}
\end{figure*}

\begin{deluxetable*}{lcccccc}
\tablenum{2}
\setlength{\tabcolsep}{4.5pt}\renewcommand{\arraystretch}{1.5}
\tablecaption{Parameters of the four subregions}
\tablewidth{0pt}
\tablehead{
\colhead{Parameters} & \colhead{} &  \colhead{G202.3+2.5} & \colhead{North tail} & \colhead{West trunk} & \colhead{South root} & \colhead{East wing} \\
}
\startdata
$\sigma_v\, {\rm (km\,s^{-1})}^a$ & (1) & $0.66\pm0.21$ & $0.48\pm0.10$ & $0.66\pm0.13$ & $0.65\pm0.20$ &  $0.80\pm0.23$\\
${\rm N}({\rm H}_2)\, (10^{22}\,{\rm cm}^{-2})^a$ & (2) & $3.18\pm0.47$ &$2.49\pm0.42$ & $4.37\pm0.67$ & $3.77\pm0.25$ & $3.08\pm0.42$\\
${\rm M}\, ({\rm M}_{\odot})$ & (3) & $302\pm28$ & $45\pm6$ & $74\pm5$ & $71\pm2$ & $62\pm5$ \\
${\rm R}\, (pc)$ & (4) & 0.371 & 0.163 & 0.156 & 0.164 & 0.170 \\
$\rho (10^{-19}\,{\rm g\,cm^{-3}})$ & (5) & $2.50\pm0.23$ & $2.57\pm0.64$ & $5.03\pm1.03$ & $3.14\pm0.78$ & $2.90\pm0.72$\\
Elongation direction 1$(^{\circ})$ & (6) & 161.8 & 23.6 & 163.5 & 148.9 & 129.8\\
Elongation direction 2$(^{\circ})$ & (7) & 162.5 & 23.9 & 166.8 & 155.3 & 166.8\\
$\theta (^{\circ})$ & (8) & $113.4$ & $88.4$ & $97.5$ & $127.3$ & $115.3$\\
$\sigma_{\theta} (^{\circ})$ & (9) & $33.0\pm3.8$ & $42.0\pm14.0$ & $34.1\pm8.8$ & $26.3\pm5.2$ & $39.9\pm9.3$\\
$(\frac{\langle B_t^{\phantom{t}2}\rangle}{\langle B_0^{\phantom{0}2}\rangle})^{1/2}$ & (10) & 0.52 & 0.59 & 0.51 & 0.40 & 0.64\\
$(\frac{\langle B_t^{\phantom{t}2}\rangle}{\langle B^2\rangle})^{1/2}$ & (11) & 0.46 & 0.51 & 0.45 & 0.37 & 0.54\\
$B_{\rm pos,adf}\, (\mu{\rm G})$ & (12) & $52.7\pm24.5$ & $35.7\pm11.6$ & $69.3\pm22.1$ & $73.0\pm28.9$ & $59.5\pm22.7$\\
$B_{\rm pos,dcf}\, (\mu{\rm G})$ & (13) & $89.6\pm36.7$ & $47.9\pm15.6$ & $110.8\pm35.3$ & $130.6\pm51.8$ & $74.2\pm28.3$\\
$\lambda_{\rm adf}$ & (14) & $3.5\pm1.8$ & $5.3\pm1.9$ & $4.8\pm1.7$ & $3.9\pm1.6$ & $3.9\pm1.6$\\
$\lambda_{\rm dcf}$ & (15) & $2.1\pm1.1$ & $4.0\pm1.5$ & $3.0\pm1.1$ & $2.2\pm0.9$ & $3.2\pm1.3$\\
$v_{A,adf} {\rm (km\,s^{-1})}$ & (16) & $0.38\pm0.16$ & $0.25\pm0.09$ & $0.39\pm0.13$ & $0.47\pm0.19$ & $0.40\pm0.16$\\
$v_{A,dcf} {\rm (km\,s^{-1})} $ & (17) & $0.64\pm0.28$ & $0.34\pm0.12$ & $0.62\pm0.21$ & $0.84\pm0.35$ & $0.49\pm0.20$\\
$\mathcal{M}_{A,adf}$ & (18) & $2.99\pm1.61$ & $3.29\pm1.34$ & $2.94\pm1.16$ & $2.41\pm1.24$ & $3.49\pm1.72$\\
$\mathcal{M}_{A,dcf}$ & (19) & $1.77\pm0.95$ & $2.45\pm0.99$ & $1.84\pm0.73$ & $1.34\pm0.70$ & $2.80\pm1.38$\\
$\alpha_{vir}$ & (20) & $0.37-0.61^b$ & $0.58-0.96$ & $0.65-1.08$ & $0.68-1.13$ & $1.23-2.05$\\
$\alpha_{vir,B,adf}$ & (21) & $0.48-0.74^b$ & $0.64-1.03$ & $0.71-1.15$ & $0.77-1.19$ & $1.28-2.10$\\
$\alpha_{vir,B,dcf}$ & (22) & $0.62-0.90^b$ & $0.68-1.03$ & $0.80-1.18$ & $0.92-1.31$ & $1.31-2.10$\\
Relative importance & (23) & $G > T > B^b$ & $G > T > B$ & $G \geq T > B$ & $G \sim T > B$ & $T > G > B$\\
\enddata
\caption{The columns of the upper table give parameters of subregions: (1) velocity dispersion; (2) mean column density; (3) average mass of masses calculated above ${\rm N}({\rm H}_2)$ level of $8.0\times10^{21}\,{\rm cm}^{-2}$ and $1.5\times10^{22}\,{\rm cm}^{-2}$; (4) effective radius; (5) mean density; (6) the elongation direction of the structure derived by auto-correlation function above column density of $8.0\times10^{21}\,{\rm cm}^{-2}$; (7) the elongation direction of the structure derived by auto-correlation function above column density of $1.5\times10^{22}\,{\rm cm}^{-2}$ ; (8) the mean B-field orientation direction; (9)the angular dispersion; (10) turbulent-to-uniform B-field strength on POS; (11) turbulent-to-total B-field strength on POS; (12) total POS B-field strength derived by ADF; (13) total POS B-field strength derived by classical DCF; (14) magnetic stability critical parameter based on $B_{\rm pos,adf}$ (not consider the factor of geometrical effect); (15) magnetic stability critical parameter based on $B_{\rm pos,dcf}$ (not consider the factor of geometrical effect); (16) Alfv\'{e}n wave velocity based on $B_{\rm pos,adf}$; (17) Alfv\'{e}n wave velocity based on $B_{\rm pos,dcf}$; (18) Alfv\'{e}n Mach number based on $B_{\rm pos,adf}$; (19) Alfv\'{e}n Mach number based on $B_{\rm pos,dcf}$ ; (20) range of virial parameter by applying the value of a from 0 to 2; (21) range of magnetic virial parameter based on $B_{\rm pos,adf}$ by applying the value of a from 0 to 2; (22) range of magnetic virial parameter based on $B_{\rm pos,dcf}$ by applying the value of a from 0 to 2; (23) Relative importance between gravity (G), B-field (B) and turbulence (T). (a) The uncertainties in (1) and (2) are the standard deviations of means; (b) The shape of G202.3+2.5 is filamentary, and the effective radius value and related parameters are for reference only.}
\label{tb_subregions}
\end{deluxetable*}

\begin{deluxetable}{lcc}
\tablenum{3}
\tablecaption{Parameters of simulated filaments}
\setlength{\tabcolsep}{15pt}
\tablewidth{5pt}
\tablehead{
}
\startdata
$N({\rm H}_2)\, (10^{22}\,{\rm cm}^{-2})$ & (1) & $3.64\pm2.47$ \\
$\sigma_v\, {\rm (km\,s^{-1})}$ & (2) & $0.44\pm0.31$ \\
$\theta (^{\circ})$ & (3) & $118.8\pm40.3$ \\
$B_{\rm pos,adf}\,(\mu{\rm G})$ & (4) & $33.9\pm24.1$ \\
$B_{\rm pos,dcf}\,(\mu{\rm G})$ & (5) & $43.3\pm30.8$ \\
$B_{\rm 3D}\,(\mu{\rm G})$ & (6) & $129.9$ \\
\enddata
\caption{The columns of the upper table give parameters of simulated filament: (1) average column density; (2) velocity dispersion; (3) the mean B-field orientation direction with the angular dispersion; (4) total POS B-field strength estimated by ADF; (5) total POS B-field strength estimated by classical DCF; (6) real density-weighted 3D B-field strength. The uncertainties in (1), (2) are the standard deviations of means.} 
\label{tb_sim}
\end{deluxetable}

\subsection{Collision filaments in simulation}\label{sec:sim}
We note the analyses above depend on the estimation of the B-field strength a lot, specifically, the reliability of the DCF and ADF method. Here we try to investigate the effects based on numerical simulation data.
\subsubsection{B-field properties in simulation}
As shown in Figure~\ref{fig:n2hp} (c), the values of fitted optical depth, $\tau$, in G202.3+2.5 are mostly smaller than 1 with a density-weighted average of 0.27, suggesting an optical thin status. \cite{Li&Klein2019} also assumed an optical thin condition, and demonstrated a collision of two magnetized filamentary clouds by an ideal MHD simulation, which resulted in a similar Y-shape configuration to G202.3+2.5 with similar column density and velocity dispersion (Table~\ref{tb_sim}). 

Figure~\ref{fig:sim} (a) shows the column density map, the collision between the two filaments disturbs the B-field in the junction region. As shown in Figure~\ref{fig:sim} (b-c)\footnote{We have chosen the upper right part of the filaments as shown in Figure~\ref{fig:sim} (b-c) to make a fair comparison since G202.3+2.5 does not have the long tail in the south and east as the simulation.}, the relative orientation between the B-field and filaments is similar to G202.3+2.5: in the east filament, the B-field shows a similar trend from parallel to perpendicular as the east wing in G202.3+2.5; in the north tail, the B-field is perpendicular to the filament as in the north tail and west trunk in G202.3+2.5; in the junction region, the B-field is compressed to be parallel with the filament like in the south root in G202.3+2.5. The distribution of the B-field orientation shown in Figure~\ref{fig:sim} (d), is close to the one in Figure~\ref{fig:b_cd} as the elongation of the gas structures in simulation and observation are both roughly in a north-south direction. The density-weighted 3-D rms B-field strength (${\rm B}_{\rm 3D}$) of the region shown in Figure~\ref{fig:sim} (b-c) is $129.9\,\mu{\rm G}$. Thus, we can apply the same methods as described in Section~\ref{sec:3.2} to investigate the reliability, specifically, the ${\rm B}_{\rm pos}$-to-${\rm B}_{\rm 3D}$ factor, or say the effect of projection. %The details of initial physical properties can be found in \citep{Li&Klein2019}. For a fair comparison, we have binned the pixel size of the simulation from $2.22\times10^{-3}$ pc to 0.05 pc. 

\subsubsection{Is the DCF or ADF reliable
towards G202.3+2.5?}
We note it is necessary to introduce the radiative transfer model if making a direct comparison between observation and simulation, but as (1) our purpose is investigating whether the DCF and ADF methods can reflect ${\rm B}_{\rm 3D}$ well towards filaments like G202.3+2.5; (2) the radiative transfer model would introduce more uncertainties and noises that affect the result \citep{Juvela2018b}, we didn't include the radiative transfer model into simulation data.

% \begin{equation}
%     Q+iU \propto \int_0^D dz\,F(z)\frac{B_x^{\phantom{x}2}-B_y^{\phantom{y}2}+2iB_x B_y}{B_x^{\phantom{y}2}+B_y^{\phantom{y}2}}
% \end{equation}

Based on the same methods as mentioned in Section~\ref{sec:3.2}, ${\rm B}_{\rm pos,adf}$ and ${\rm B}_{\rm pos,dcf}$ of the region shown in Figure~\ref{fig:sim} (b-c) are estimated as $\sim34\, \mu \rm{G}$ and $\sim43\, \mu \rm{G}$, respectively (Table~\ref{tb_sim}). And the real density-weighted 3D B-field strength ($B_{\rm 3D}$) is $129.9\,\mu{\rm G}$, suggesting a factor of $\sim3-4$ is needed to estimate a ${\rm B}_{\rm 3D}$ from ${\rm B}_{\rm pos}$ derived from DCF or ADF method. While \cite{Crutcher2004} assumed a random projection of ${\rm B}_{\rm 3D}$ to generate a factor of $4/\pi$ statistically, \cite{LiuJH2021} concluded a factor of $\sqrt{3/2}$ based on different simulation results, both of them are much smaller than 3 in our case. There are three possible reasons: 

(1) the factors $4/\pi$ and $\sqrt{3/2}$ are all derived statistically. In a specific case like this work, they may have a large bias; 

(2) ADF and DCF methods may underestimate ${\rm B}_{\rm pos}$ in this case. \cite{LiuJH2021} simulated self-gravitating clustered massive star-forming regions at scales no larger than 0.5 pc, and suggested a $f_{\rm DCF}$ of 0.21 in ADF method, which shows 45\% uncertainty at scales between 0.1 and 1 pc. While filaments in our case have lengths longer than 1 pc, the factor of 0.21 may not be suitable here. If applied the original $f_{\rm DCF}$, 0.5, ${\rm B}_{\rm pos,adf}\,(\mu{\rm G})$ would be $\sim81\,\mu{\rm G}$, and the ratio between ${\rm B}_{\rm 3D}$ and ${\rm B}_{\rm pos}$ would be closer to $4/\pi$ and $\sqrt{3/2}$, this suggests that in regions with scale $\geq$ 1 pc or gravity hasn't dominated, a larger $f_{\rm DCF}$ will be more convincing.

(3) Both (1) and (2) contribute to this result. It is necessary to introduce MHD simulation results when estimating ${\rm B}_{\rm 3D}$ from observations, as specific cases may need specific factors due to the projection effect.

Based on the simulation result, a new estimation of ${\rm B}_{\rm 3D}$ in G202.3+2.5, is given as ${\rm B}_{\rm 3D, adf}\sim202\,\mu{\rm G}$ and ${\rm B}_{\rm 3D, dcf}\sim269\,\mu{\rm G}$, respectively. The three-times-larger ${\rm B}_{\rm 3D}$ would enlarge ${\rm M}_{\rm B}$ and $v_A$, and further, increase $\alpha_{\rm vir}$ by a factor of $\sim1.5-2$ and decrease $\mathcal{M}_{A}$ by a factor of $\sim3$. Thus, the relative importance of the B-field is probably underestimated in Section~\ref{sec:GBT}, which is possibly as comparable as gravity and turbulence in the west trunk and south root.

\subsection{Uncertainties Analysis}
One of the main challenges in the quantitative analysis discussed above comes from the uncertainties that result from observations and parameter estimation. 

\begin{figure*}[ht!]
\centering\includegraphics[scale=0.7]{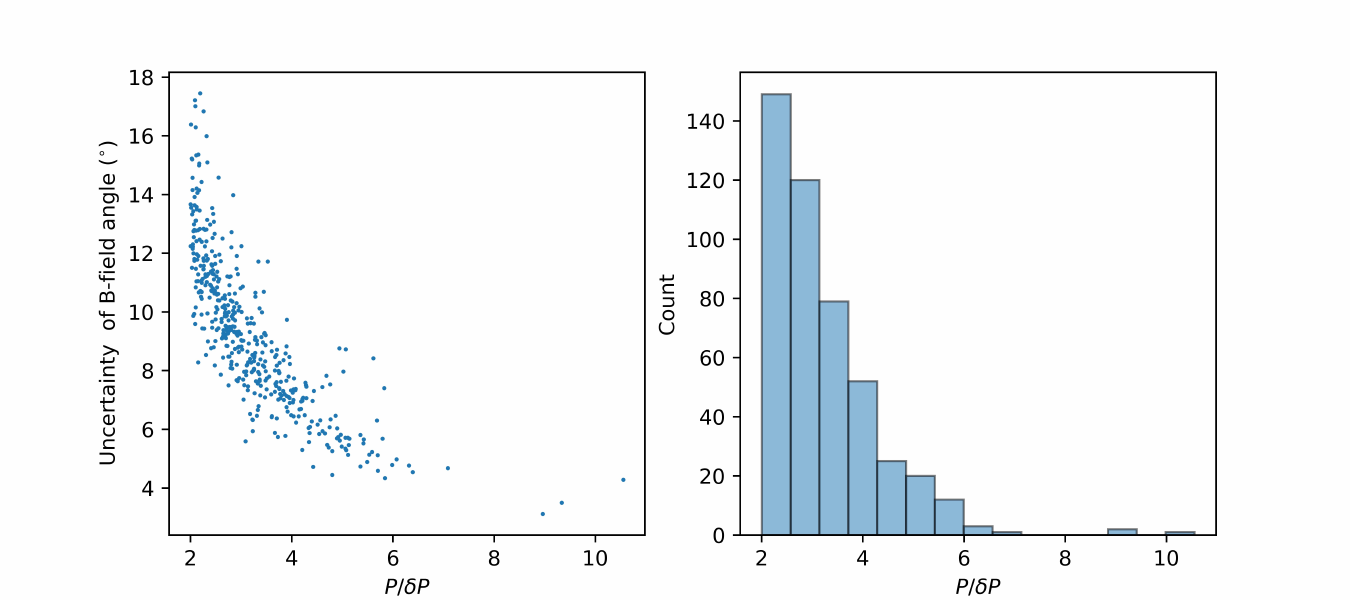}
\caption{\em{\textbf{Left panel}: Correlation between uncertainty of B-field angles from JCMT/POL-2 observation and corresponding S/N of polarization intensity, $P/\delta P$. \textbf{Right panel}: Distribution of $P/\delta P$ level of JCMT/POL-2 B-field data. %angles.
}}
\label{fig:snr_db}
\end{figure*}

(1) The polarization angle. Due to the difficulty of ground-based polarization observations, although we observed for more than 20 hrs, the data quality is still not so high. As shown in Figure~\ref{fig:snr_db}, $\sim66.8\%$ of the JCMT/POL-2 pseudo-vectors have an angle uncertainty larger than $8^{\circ}$ and $\sim81.9\%$ have an S/N below 4. Also, the intrinsic large dispersion of the polarization angles will affect the calculation of ADF as well as the ratio of turbulent-to-uniform field strength, $(\frac{\langle B_t^{\phantom{t}2}\rangle}{\langle B_0^{\phantom{0}2}\rangle})^{-1/2}$, 

(2) LOS depth of the structures. We calculate the volume of subregions by multiplying the area that is above the $N({\rm H}_2)$ contour of $1.5\times10^{22}\,{\rm cm}^{-2}$ and the estimated depth with a value of 0.2\,pc based on the cylindrical shape. Here, an uncertainty of 0.05\,pc is expected, which would cause a $\sim25\%$ uncertainty in the density, causing a$\sim15\%$ uncertainty in the calculation of B-field strength due to the error propagation.

(3) The correction factor, $f_{\rm DCF}$, in Equation~\ref{eq:dcf}. The ADF and classical DCF take different $f_{\rm DCF}$, which are 0.21 and 0.5, respectively. Classical $f_{\rm DCF}$ is taken as 0.5 based on simulation result from \cite{Ostriker2001}, which is applicable on low-density regions with scales larger than 1 pc and $\sigma_{\theta} < 25^{\circ}$, thus $\tan(\sigma_{\theta})$ is applied to replace $\sigma_{\theta}$ to remove the small $\sigma_{\theta}$ restriction. Later, \cite{LiuJH2021} adopted the value of 0.21, as they found the contribution from the ordered field structure could cause an overestimation of the angular dispersion by an average factor of $\sim2.5$ when calibrating the ADF method. The value of $f_{\rm DCF}$ affects the final result of B-field strength directly, and as shown in the result in Section~\ref{sec:sim}, both methods may underestimate the B-field strength by a factor of $\sim2-3$ but still within an acceptable range. Thus, we suggest it is necessary to compare with similar simulation cases to constrain the result.

(4) The estimation of $B_{\rm 3D}$. \cite{Crutcher2004} suggested $B_{\rm 3D} = \frac{4}{\pi}B_{pos}$ statistically, which is applicable for the big sample where \textbf{$B_{\rm 3D}$} shows a random alignment with the POS. For a specific case like G202.3+2.5, the factor of $\frac{4}{\pi}$ would be a very rough estimation by comparing the simulation result shown in Section~\ref{sec:sim}, which gives a factor of $\sim3-4$. And this further affects the calculation of Alfv\'{e}n wave velocity ($v_A$), Alfv\'{e}n Mach number ($\mathcal{M}_A$) and the magnetic virial parameter ($\alpha_{vir,B}$). 

To sum up, the uncertainties are not negligible in the quantitative analysis that is related to B-field strength, and we suggest that only when these parameters are much larger or smaller than the critical value (e.g. $\lambda\geq10$, $\alpha_{vir,B} \ll 1$), the conclusions judging the critical status could be confirmed.

\section{Summary}\label{sec5}
In this work, we have taken 850 $\mu$m continuum polarization observation towards a nearby star-forming cloud, G202.3+2.5, using the JCMT/POL-2 to investigate the B-field morphology. Combining these with ${\rm N_2H^{+}}$ observations using IRAM, we also estimate the B-field strength via two DCF methods and study the relative importance of gravity (G), B-field (B), and turbulence (T). Our conclusions are as follows:

(1) Planck data show a parallelism alignment between the B-field and the host filament of G202.3+2.5 at a larger scale ($\sim$1\,pc). Based on JCMT/POL-2, we find the mean B-field on POS across G202.3+2.5 is roughly perpendicular to the whole cloud at a scale of 0.05\,pc, showing a typical parallel-to-perpendicular trend found in \cite{PlanckXXXV}. Specifically, The B-field behaves differently in subregions of G202.3+2.5: in the north tail and the west trunk is perpendicular to the filament structure, while in the east wing, it is disturbed by the collision, showing a large angular dispersion, and is compressed to be parallel with the main filament and gas flow in the south root, showing a similar perpendicular-to-parallel transition in high-density region to Serpens South Cloud\citep{Pillai2020}. The compressed signature of the B-field and perpendicular-to-parallel transition could be a hint or evidence of the gas flow in the high-density part of a star formation region. 

(2) The POS B-field strengths of the four subregions derived from the ADF method are between 36 and 73 $\mu$G, with an uncertainty level of $\sim30\%$, and the ones derived from classical DCF are $\sim30-80\%$ higher. The Alfv\'{e}n Mach number, magnetic stability critical parameter, and magnetic virial parameter are then derived to estimate the critical status. We conclude $G > T > B$ in the north tail, $G \geq T > B$ in the west trunk, $G\sim T > B$ in the south root, and $T > G > B$ in the east wing. 

(3) We compare our observation with the simulation of \cite{Li&Klein2019}, finding the ratio between real ${\rm B}_{\rm 3D}$ and ${\rm B}_{\rm pos}$ is $\sim3-4$, which is much larger than the widely-used values, $4/\pi$ \citep{Crutcher2004} and $\sqrt{3/2}$ \citep{LiuJH2021}. We suppose the reasons might be: (a) classical DCF and ADF methods would underestimate the POS B-field strength in filaments-colliding regions like G202.3+2.5, a $f_{\rm DCF}$ of 0.5 is more convincing than 0.21 in regions with a scale larger than 1 pc; (b) the ${\rm B}_{\rm pos}$-to-${\rm B}_{\rm 3D}$ factor of $4/\pi$ is not suitable for some specific case like this work. Thus, the relative importance of B-field is probably underestimated and it is necessary to compare with the simulation to constrain the result.

\section*{Acknowledgements}
This work has been supported by the National Key R\&D Program of China (No. 2022YFA1603100). This work has been supported by Shanghai Rising-Star Program (23YF1455600). T.L. acknowledges supports from the National Natural Science Foundation of China (NSFC), through grants No. 12073061 and No. 12122307; the international partnership program of the Chinese Academy of Sciences, through grant No. 114231KYSB20200009; and the Shanghai Pujiang Program 20PJ1415500. J.L. is partially supported by a Grant-in-Aid for Scientific Research (KAKENHI Number JP23H01221) of JSPS. We thank Hauyu Baobab Liu for helpful comments and discussions. MJ acknowledges support from the Research Council of Finland grant No. 348342. The work of MGR is supported by NOIRLab, which is managed by the Association of Universities for Research in Astronomy (AURA) under a cooperative agreement with the National Science Foundation. K.T. was supported by JSPS KAKENHI (Grant Number JP20H05645).

The James Clerk Maxwell Telescope is operated by the East Asian Observatory on behalf of The National Astronomical Observatory of Japan; Academia Sinica Institute of Astronomy and Astrophysics; the Korea Astronomy and Space Science Institute; the National Astronomical Research Institute of Thailand; Center for Astronomical Mega-Science (as well as the National Key R\&D Program of China with No. 2017YFA0402700). Additional funding support is provided by the Science and Technology Facilities Council of the United Kingdom and participating universities and organizations in the United Kingdom and Canada. Additional funds for the construction of SCUBA-2 were provided by the Canada Foundation for Innovation.

\software{Starlink \citep{berry2005,Chapin2013}, Astropy \citep{astropy2013}}
% All of these commands work equally well in text and math mode.

% \section{Using Chinese, Japanese, and Korean characters}

% Authors have the option to include names in Chinese, Japanese, or Korean (CJK) 
% characters in addition to the English name. The names will be displayed 
% in parentheses after the English name. The way to do this in AASTeX is to 
% use the CJK package available at \url{https://ctan.org/pkg/cjk?lang=en}.
% Further details on how to implement this and solutions for common problems,
% please go to \url{https://journals.aas.org/nonroman/}.


\bibliography{sample631.bbl}{}
\bibliographystyle{aasjournal}

% \section*{Appendix}

% \subsection*{A. The elongation direction of filament structures}\label{sec:ac}


\end{CJK}
\end{document}